\documentclass[11pt,a4paper]{article}
\usepackage[T1]{fontenc}
\usepackage[utf8]{inputenc}
\usepackage{lmodern}
\usepackage[a4paper,margin=1in,footskip=24pt]{geometry}
\usepackage{amsmath,amssymb,amsthm}
\usepackage{graphicx}
\usepackage{booktabs,array}
\usepackage{setspace}
\usepackage{microtype}
\usepackage[authoryear,round]{natbib}
\bibpunct[, ]{(}{)}{,}{a}{}{,}
\usepackage[font=small,labelfont=bf,labelsep=period]{caption}
\usepackage{etoolbox}
\robustify{\footnotesize}
\robustify{\small}
\usepackage[colorlinks=true,linkcolor=blue,citecolor=blue,urlcolor=blue]{hyperref}
\usepackage{refcount}
\hypersetup{
  pdftitle={A Rough Instantaneous Squared VIX Model for Term Structure and Option Pricing},
  pdfauthor={Tianfeng Yang; Wen Su; Blanka Horvath},
  pdfkeywords={VIX; VIX term structure; VIX options; Affine Volterra model}
}

\theoremstyle{plain}
\newtheorem{theorem}{Theorem}[section]
\newtheorem{corollary}{Corollary}[section]
\newtheorem{lemma}{Lemma}[section]
\newtheorem{proposition}{Proposition}[section]
\theoremstyle{definition}

\theoremstyle{remark}

\newcommand{\appref}[1]{\hyperref[#1]{Appendix~\getrefnumber{#1}}}

\begin{document}

\begingroup
\renewcommand{\thefootnote}{\fnsymbol{footnote}}
\begin{center}
{\setstretch{1.1}\LARGE\bfseries
A Rough Instantaneous Squared VIX Model\\[3pt]
for Term Structure and Option Pricing\par}
\vspace{14pt}
{\large Tianfeng Yang$^{\dagger}$, Wen Su\footnotemark[1]$^{\ddagger}$, Blanka Horvath$^{\ddagger}$\par}
\vspace{9pt}
{\small\setstretch{1.1}
$^{\dagger}$Department of Mathematical Sciences, Tsinghua University, China\\
$^{\ddagger}$Mathematical Institute, University of Oxford, OX2 6GG, UK\par}
\end{center}
\footnotetext[1]{Corresponding author. Email: \href{mailto:wen.su@maths.ox.ac.uk}{wen.su@maths.ox.ac.uk}}
\endgroup
\vspace{4pt}

\begin{abstract}
\setstretch{1}

The VIX term structure provides forward-looking measures of market uncertainty across horizons, while VIX options allow investors to hedge volatility risk.
This paper proposes a rough instantaneous squared VIX (RISVIX) model by introducing fractional memory into the conventional Markovian framework.
We derive an explicit representation of the VIX term structure governed by Mittag--Leffler decay and a semi-closed-form pricing formula for VIX options based on an affine Fourier--Laplace transform.
These results support a two-step calibration procedure that estimates memory, mean reversion, and date-specific initial states from VIX term-structure indices before calibrating the remaining distributional parameters to VIX option prices.
Using daily VIX term-structure indices and VIX option quotes from January 2018 to June 2025, we find that RISVIX improves upon the Markovian benchmark in fitting the VIX term structure, with larger gains during elevated-volatility periods and for its slope and curvature.
RISVIX also improves out-of-sample option pricing, reducing MSE and MAE by 13.06\% and 6.33\%, respectively, with larger gains at shorter maturities and in high-VIX states.
These findings suggest that the gains from fractional memory are particularly pronounced when volatility is elevated and can be obtained within a tractable direct model of VIX derivatives. Comparisons with additional benchmarks from the recent literature also show improvements to varying degrees in both VIX term-structure fitting and option pricing.

\end{abstract}
\vspace{5pt}

\vspace{8pt}

\section{Introduction}
\label{sec1}

Since the global financial crisis of 2008, the VIX has become a central gauge of financial-market uncertainty and the reference index for a major market in volatility derivatives \citep{whaley2000fear,LZZ2019}.
The index itself is not directly tradable: it is calculated from S\&P 500 option prices and summarizes the market's risk-neutral expectation of S\&P 500 variation over the next 30 days \citep{LZ2012}.
Tradable exposure is instead provided by VIX futures and options, which allow investors to hedge volatility risk and take directional positions on changes in market uncertainty \citep{LZZ2019}. 
At the same time, evidence that volatility exhibits rough, non-Markovian dynamics has prompted growing interest in the implications of roughness for VIX futures and options \citep{GJR2018,JacquierMartiniMuguruza2018,Romer2022}. 
The prominence of the VIX market, together with this evidence, creates a need for models that can capture the dynamics of the VIX while remaining sufficiently tractable for derivative pricing.

Existing models for VIX derivatives can be organized into two broad approaches. 
The first derives VIX from the joint risk-neutral dynamics of the S\&P 500 and its latent variance process, thereby maintaining a structural link between SPX and VIX derivatives, but often requiring a relatively rich specification and joint calibration to both markets \citep{ZZ2006,ZZ2007,CK2013,LLLC2017,BGL2019,EY2022}.
The second approach models variance, volatility, VIX, or VIX futures directly \citep{Lin2007,DDC2011,LZ2012,MS2013,HN2013,LZZ2019}. 
By avoiding an explicit model for the SPX spot price, direct specifications are generally more parsimonious and yield more tractable VIX derivative prices, although they do not impose a structural link between the equity and volatility markets.

This paper follows the instantaneous squared VIX (ISVIX) approach developed by \citet{LZ2012,LZZ2019}, which models directly the short-horizon limiting object whose conditional average determines squared VIX.
The resulting Markovian specification is parsimonious and tractable, but it restricts the transmission of current volatility conditions across maturities to the pattern implied by exponential mean reversion.
A natural way to relax this restriction is to introduce memory. 
Rough-volatility models use singular fractional kernels to capture the low local regularity and history dependence observed in volatility dynamics \citep{BFG2016,GJR2018,ER2019}.
Most applications to VIX derivatives, however, introduce roughness into the latent variance or forward-variance process of the equity index and then derive VIX as a nonlinear functional of that process. 
This route is well suited to joint SPX--VIX modeling, but it makes the VIX term structure and VIX option prices depend on a comparatively rich latent-state system. 

We instead propose a rough instantaneous squared VIX (RISVIX) model, by introducing a fractional Volterra kernel directly into the ISVIX framework of \citet{LZZ2019}.
We derive an explicit VIX term structure governed by a Mittag--Leffler decay function and a semi-closed pricing formula for VIX options based on an affine Fourier--Laplace transform.
These results lead to a two-step procedure in which the VIX term structure is used to estimate the memory and mean-reversion parameters, while VIX options are used to calibrate the remaining distributional parameters.
Empirically, we estimate RISVIX using VIX term-structure indices and VIX option prices from 2018 to 2025.
The gains from fractional memory vary substantially with market conditions.
RISVIX improves the fit of the VIX term structure, with larger gains for its slope and curvature and during periods of elevated volatility.
It also improves out-of-sample VIX option pricing, reducing MSE and MAE by 13.06\% and 6.33\%, respectively, with larger gains at shorter maturities and in high-VIX states.

Our contributions are threefold. 
First, we introduce fractional memory into an ISVIX model, yielding a parsimonious rough framework for modeling the VIX term structure and pricing VIX options. 
Second, we retain the tractability of direct VIX modeling by deriving an explicit term-structure formula and a semi-closed-form option-pricing formula, which together support a transparent two-step estimation and calibration procedure. 
Third, we show empirically that the gains from fractional memory vary substantially across volatility regimes: RISVIX improves the fit of the VIX term structure and the out-of-sample pricing of VIX options more strongly when volatility is elevated.

\vspace{1em}
\noindent \textbf{Related literature.}
Our paper is related to two strands of literature.
First, it contributes to the literature on direct modeling of VIX term structures and derivatives. 
\citet{Lin2007} price VIX futures by integrating information under the physical and risk-neutral measures, while \citet{DDC2011} develop a direct volatility-futures model and compare CEV and CIR specifications with and without jumps. 
\citet{HN2013} model the term structure of VIX futures directly rather than deriving it from an equity-variance process, and \citet{MS2013} conduct an extensive empirical evaluation of VIX derivative valuation models across different market conditions. 
More closely related to our work, \citet{LZ2012} extend VIX to arbitrary horizons and introduce instantaneous and long-term squared VIX as the short- and long-maturity limits of the VIX term structure. Building on this representation, \citet{LZZ2019} develop a unified Markovian ISVIX model with a stochastic mean level and jumps for pricing VIX derivatives across horizons.
RISVIX extends their framework by replacing the propagation pattern implied by Markovian mean reversion with a fractional one.

Second, our paper contributes to the literature on rough volatility and VIX derivatives.
Rough-volatility models use fractional kernels to capture the low local regularity and history dependence of volatility \citep{BFG2016,GJR2018,ER2019}.
Existing applications to VIX derivatives generally specify rough dynamics for the latent variance or forward-variance curve of the equity index and then derive VIX from that process \citep{JacquierMartiniMuguruza2018,HJT2020,Romer2022,BourgeyDeMarcoGobet2023,JacquierMuguruzaPannier2025}.
The closest study is \citet{YeFanKwok2026}, who derive the VIX term structure from rough Heston using a Markovian approximation. 
Our RISVIX differs from their approach in two respects. 
First, it retains the fractional Volterra kernel directly in the ISVIX dynamics, thereby preserving the non-Markovian memory structure without relying on a Markovian approximation.
Second, its affine Volterra structure yields a semi-closed pricing formula for VIX options, allowing the VIX term structure and option prices to be studied within the same framework.

\vspace{1em}
The remainder of the paper is organized as follows. \autoref{sec2} reviews the ISVIX framework, fractional Brownian motion and affine Volterra processes.
\autoref{sec3} introduces RISVIX, derives the VIX term structure and option-pricing formula, and presents the estimation procedure.
\autoref{sec4} reports the empirical results for the VIX term structure and VIX options. 
\autoref{sec5} concludes.
The Appendix contains the proofs and auxiliary results.

\section{Preliminaries}
\label{sec2}

This section reviews instantaneous squared VIX, fractional Brownian motion and affine Volterra models, which will be used throughout this paper.
We work on a filtered probability space $(\Omega,\mathcal{F},(\mathcal{F}_t)_{t\geq0},\mathbb{Q})$ satisfying the usual conditions, where $\mathbb{Q}$ is a risk-neutral measure.
Write $\mathbb{E}_t^{\mathbb{Q}}[\cdot]=\mathbb{E}^{\mathbb{Q}}[\cdot\mid\mathcal{F}_t]$.

\subsection{Instantaneous Squared VIX}
\label{subsec:isvix}

Let $S_t$ be a strictly positive semimartingale representing the equity index.
We use the continuous-strike idealisation of VIX, whose squared value over a horizon $\tau>0$ is represented by the log contract \citep{Lin2007,LZ2012}:
\begin{equation}
\label{eq:prelim_vix_log_contract}
\mathrm{VIX}_{t,\tau}^{2}
=
\frac{2}{\tau}\mathbb{E}_t^{\mathbb{Q}}
\left[
\int_t^{t+\tau}\frac{\mathrm{d}S_s}{S_{s-}}
-
\log\left(\frac{S_{t+\tau}}{S_t}\right)
\right].
\end{equation}
Suppose that the continuous component of the index return has instantaneous variance $v_t$, while log-price jumps have intensity $\lambda_t^x$ and sizes distributed as a random variable $x$ with a state-independent law.
When the integrated variance and jump correction have finite expectations, \autoref{eq:prelim_vix_log_contract} admits the instantaneous squared VIX (ISVIX) representation of \citet{LZ2012}:
\begin{equation}
\label{eq:prelim_isvix}
\begin{aligned}
V_t
=
v_t+2\lambda_t^x\mathbb{E}^{\mathbb{Q}}\left[e^x-1-x\right],\quad
\mathrm{VIX}_{t,\tau}^{2}
=
\frac{1}{\tau}\mathbb{E}_t^{\mathbb{Q}}
\left[\int_t^{t+\tau}V_s\,\mathrm{d}s\right].
\end{aligned}
\end{equation}
If $V$ is right-continuous in $L^1(\mathbb{Q})$, then $\mathrm{VIX}_{t,\tau}^{2}\to V_t$ in $L^1(\mathbb{Q})$ as $\tau\downarrow0$.
ISVIX coincides with the instantaneous diffusive variance in the absence of price jumps and otherwise includes a jump correction \citep{LZZ2019}.

\subsection{Fractional Brownian Motion and Fractional Kernels}
\label{subsec:fbm}

A standard fractional Brownian motion (fBM) with Hurst parameter $H\in(0,1)$ is a centred Gaussian process $B^H$ with covariance \citep{MVN1968}
\begin{equation}
\label{eq:prelim_fbm_covariance}
\mathbb{E}^{\mathbb{Q}}\left[B_s^H B_t^H\right]
=
\frac{1}{2}\left(s^{2H}+t^{2H}-|t-s|^{2H}\right),
\qquad s,t\geq0.
\end{equation}
It has stationary increments and admits sample paths that are locally H\"older continuous of every order strictly less than $H$.
The case $H=1/2$ is standard Brownian motion, whereas $H<1/2$ gives rougher paths.

We restrict attention to $H\in(0,1/2]$ and set $\alpha=H+1/2$.
The corresponding fractional kernel is $K(t)=t^{\alpha-1}/\Gamma(\alpha)$ for $t>0$, where $\Gamma$ is the gamma function.
For a standard Brownian motion $B$, define
\begin{equation}
\label{eq:prelim_riemann_liouville}
\widetilde{B}_t^H
=
\int_0^t K(t-s)\,\mathrm{d}B_s.
\end{equation}
For $H<1/2$, this Riemann--Liouville process differs from standard fBM: its increments are not stationary, although its paths have the same local H\"older regularity \citep{BFG2016}.
Since $K$ is square-integrable on every bounded interval for $\alpha>1/2$, the convolution is a Brownian It\^o integral.
At $\alpha=1$, $K\equiv1$ and $\widetilde{B}^{1/2}=B$.

\subsection{Affine Volterra Models}
\label{sec:AffineVolterra}

We consider a class of affine Volterra models with diffusion and jump innovations \citep{ALP2019,BLP2022}:
\begin{equation}
\label{eq:prelim_affine_volterra}
X_t
=
X_0
+
\int_0^t
\mathcal{K}(t-s)\circ
\left\{
\mu(X_s)\,\mathrm{d}s
+
\sigma(X_s)\,\mathrm{d}B_s
+
\mathrm{d}J_s
\right\},
\end{equation}
where $X$ is a predictable $d$-dimensional process with locally integrable paths, $X_0$ is deterministic, $B$ is an $m$-dimensional Brownian motion, $\mathcal{K}=(K_1,\ldots,K_d)^\top$ consists of deterministic, locally square-integrable kernels, and $\circ$ denotes componentwise multiplication. Here $J$ is a compensated jump process whose jumps have conditional intensity $\lambda^y(X_s)$ and sizes distributed as a random vector $y\in\mathbb{R}^d$ with a fixed, state-independent law satisfying $\mathbb{E}^{\mathbb{Q}}[\|y\|^2]<\infty$; its compensator subtracts $\lambda^y(X_s)\mathbb{E}^{\mathbb{Q}}[y]\,\mathrm{d}s$ from the uncompensated jump increments. For singular kernels, \autoref{eq:prelim_affine_volterra} is understood to hold $\mathbb{Q}\otimes\mathrm{d}t$-almost everywhere \citep{BLP2022}.

The drift, diffusion covariance and jump intensity are affine on the state space:
\begin{equation}
\label{eq:prelim_affine_coefficients}
\begin{aligned}
\mu(x)
=
b_0+\sum_{k=1}^d b_kx_k,\,\,\,
\sigma(x)\sigma(x)^\top
=
H_0+\sum_{k=1}^d H_kx_k,\,\,\,
\lambda^y(x)
=
\lambda_0^y+\sum_{k=1}^d\lambda_k^y x_k,
\end{aligned}
\end{equation}
where $b_k\in\mathbb{R}^d$, each $H_k\in\mathbb{R}^{d\times d}$ is symmetric, and $\lambda_k^y\in\mathbb{R}$ for $k=0,\ldots,d$. We assume that the diffusion covariance is positive semidefinite and the jump intensity is nonnegative on the state space. Taking $K_i\equiv1$ for every component recovers an affine jump-diffusion, while setting $\lambda^y\equiv0$ gives the diffusion setting of \citet{ALP2019}.

Whenever the expectation is absolutely convergent, write $\varrho(z)=\mathbb{E}^{\mathbb{Q}}[e^{z^\top y}]$ and define the affine cumulant functions by
\begin{equation}
\label{eq:prelim_affine_cumulants}
R_k(z)
=
b_k^\top z
+
\frac{1}{2}z^\top H_kz
+
\lambda_k^y
\left[
\varrho(z)-1-z^\top\mathbb{E}^{\mathbb{Q}}[y]
\right],
\qquad k=0,\ldots,d.
\end{equation}
For a horizon $T>0$ and transform argument $u\in\mathbb{C}^d$, the associated Riccati--Volterra equation is
\begin{equation}
\label{eq:prelim_riccati_volterra}
\Psi(t,u)
=
u\circ\mathcal{K}(t)
+
\int_0^t
\mathcal{K}(t-s)\circ R\bigl(\Psi(s,u)\bigr)\,\mathrm{d}s,
\qquad 0<t\leq T,
\end{equation}
where $R(z)=(R_1(z),\ldots,R_d(z))^\top$.

Under suitable integrability and martingale conditions, solutions of \autoref{eq:prelim_riccati_volterra} yield initial-time exponential-affine transforms; see \citet{ALP2019,BLP2022}.
The functions $R_1,\ldots,R_d$ determine the Riccati--Volterra equation, whereas $R_0$ enters only the state-independent part of the transform exponent.
At later valuation times, conditional transforms generally depend on the past through an adjusted forward curve rather than on $X_t$ alone\citep{ALP2019,BLP2022}.

\section{Framework}
\label{sec3}

In this section, we propose a rough ISVIX (RISVIX) model and derive the corresponding VIX term structure and VIX option pricing formulas.
Throughout the theoretical analysis, time zero denotes the valuation origin.

\subsection{Model Specification}
\label{subsec:model_specification}

We extend the ISVIX framework of \citet{LZZ2019} using the fractional kernel $K(t)=t^{\alpha-1}/\Gamma(\alpha)$ introduced in \autoref{subsec:fbm}. 
Fix a finite horizon $T>0$. 
Under the risk-neutral measure $\mathbb Q$, the RISVIX model is
\begin{equation}
\label{eq:rough_isvix_model}
\begin{aligned}
V_t
&=
V_0+\int_0^t K(t-s)
\left\{
\kappa(\theta_s-V_s)\,\mathrm{d}s
+\sigma_V\sqrt{V_s}\,\mathrm{d}B_{1,s}
+\mathrm{d}J_s^y
\right\},\\
\theta_t
&=
\theta_0+\sigma_\theta\int_0^t\sqrt{\theta_s}\,\mathrm{d}B_{2,s},
\qquad 0\leq t\leq T.
\end{aligned}
\end{equation}
Here $V_0,\theta_0\geq0$ are $\mathcal F_0$-measurable initial states satisfying $\mathbb E^{\mathbb Q}[V_0^2+\theta_0^2]<\infty$, and $B_1$ and $B_2$ are independent Brownian motions. Let $N^y$ be a Poisson process with intensity $\lambda^y$, and let $\{Y_n\}_{n\geq1}$ be an independent sequence of exponentially distributed random variables with mean $\mu_y$. We assume that $B_1$, $B_2$, $N^y$ and $\{Y_n\}_{n\geq1}$ are mutually independent and independent of $\mathcal F_0$, and define $J_t^y=\sum_{n=1}^{N_t^y}Y_n-\lambda^y\mu_y t$. Thus, $J^y$ is a square-integrable martingale, and the jump-size moment generating function is $\varrho(z)=(1-\mu_y z)^{-1}$ for $\operatorname{Re}(z)<\mu_y^{-1}$.

The parameters satisfy $\kappa,\mu_y>0$, $\sigma_V,\sigma_\theta,\lambda^y\geq0$ and $\alpha\in(1/2,1)$. The parameter $\kappa$ governs mean reversion towards the stochastic level $\theta$, while $\sigma_V$ and $\sigma_\theta$ govern innovations to ISVIX and its mean level, respectively. The fractional kernel introduces memory into the ISVIX dynamics; formally setting $\alpha=1$ gives $K\equiv1$ and a Markovian jump-diffusion.

On $[0,T]$, we assume that \autoref{eq:rough_isvix_model} admits a nonnegative weak solution $(V,\theta)$. 
The process $\theta$ is continuous and nonnegative, while $V$ is predictable with $\int_0^T V_s\,\mathrm{d}s<\infty$ almost surely; both $V\geq0$ and the first equation in \autoref{eq:rough_isvix_model} hold $\mathbb Q\otimes\mathrm{d}t$-almost everywhere. 
Under the initial second-moment condition above, \autoref{lem:risvix_integrated_moments} gives $\mathbb E^{\mathbb Q}[\int_0^T(V_t^2+\theta_t^2)\,\mathrm{d}t]<\infty$ and establishes that $\theta$ is a square-integrable martingale.
General weak-existence results for affine stochastic Volterra equations can be found in \citet{AJ2021}. 
The following proposition gives the conditional mean of ISVIX.

\begin{proposition}
\label{thm:isvix_conditional_mean}
Suppose that \autoref{eq:rough_isvix_model} admits the finite-horizon solution specified in \appref{app:risvix_admissibility} on $[0,T]$. Then, the following identity holds almost surely:
\begin{equation}
\label{eq:isvix_conditional_mean}
\mathbb E_0^{\mathbb Q}[V_s]
=
\theta_0
+
(V_0-\theta_0)E_{\alpha,1}(-\kappa s^\alpha),
\end{equation}
where $E_{a,b}(z)=\sum_{n=0}^{\infty}z^n/\Gamma(an+b)$ is the two-parameter Mittag--Leffler function. 
The right-hand side gives the unique continuous representative of the conditional mean on $[0,T]$.
\end{proposition}

\noindent\textbf{Remark}.
The conditional mean in \autoref{eq:isvix_conditional_mean} depends only on $(\alpha,\kappa,V_0,\theta_0)$. The Brownian and compensated jump terms have zero conditional mean, while $\mathbb E_0^{\mathbb Q}[\theta_s]=\theta_0$. Hence, although $(\sigma_V,\sigma_\theta,\lambda^y,\mu_y)$ do not enter the first moment, they shape the conditional distribution of future ISVIX and therefore affect VIX option prices.

\subsection{VIX Term Structure}
\label{subsec:vix_term_structure}

Recall from \autoref{eq:prelim_isvix} that, for $t\geq0$ and $\tau>0$, the squared VIX is
$\mathrm{VIX}_{t,\tau}^{2}
=
\tau^{-1}\mathbb{E}_t^{\mathbb{Q}}
[\int_t^{t+\tau}V_s\,\mathrm{d}s]$.
The following theorem gives its representation under the RISVIX model.

\begin{theorem}
\label{thm:vix_term_structure}
Suppose that \autoref{eq:rough_isvix_model} admits the finite-horizon solution specified in \appref{app:risvix_admissibility} on $[0,T]$.
Then, for each fixed $t\in[0,T)$ and $\tau\in(0,T-t]$, the following identity holds almost surely:
\begin{equation}
\label{eq:vix_term_structure}
\begin{aligned}
\mathrm{VIX}_{t,\tau}^{2}
={}&
w(\tau)V_0
+
[1-w(\tau)]\theta_t\\
&+
\int_0^t
a_\tau(t-s)
\left\{
\kappa(\theta_s-V_0)\,\mathrm{d}s
+
\sigma_V\sqrt{V_s}\,\mathrm{d}B_{1,s}
+
\mathrm{d}J_s^y
\right\},
\end{aligned}
\end{equation}
where $w(\tau)=E_{\alpha,2}(-\kappa\tau^\alpha)$ and
$a_\tau(s)=
[\,
E_{\alpha,1}(-\kappa s^\alpha)
-
E_{\alpha,1}(-\kappa(s+\tau)^\alpha)
\,]/(\kappa\tau)$.
\end{theorem}

\noindent\textbf{Remark}.
At the valuation origin, \autoref{eq:vix_term_structure} reduces to
$\mathrm{VIX}_{0,\tau}^{2}
=
V_0w(\tau)+\theta_0[1-w(\tau)]$.
The weight $w(\tau)$ decreases from one to zero as $\tau$ increases.
Hence, over any horizon on which the finite-horizon solution is defined, the time-zero squared VIX curve moves monotonically from $V_0$ towards $\theta_0$: it is upward sloping when $V_0<\theta_0$, downward sloping when $V_0>\theta_0$, and flat when the two levels coincide.
The time-zero term structure depends only on $(\alpha,\kappa,V_0,\theta_0)$.
The diffusion and jump parameters do not enter this curve, but they affect the distribution of future VIX through the history term in \autoref{eq:vix_term_structure}.
This separation motivates the two-stage calibration used below.

\subsection{VIX Option Pricing}
\label{subsec:vix_option_pricing}

Consider a European VIX call valued at time zero, with expiry $T>0$, strike $K>0$, and variance horizon $\tau_0>0$.
By \autoref{thm:vix_term_structure}, its terminal squared VIX is given by \autoref{eq:vix_term_structure} evaluated at $t=T$ and $\tau=\tau_0$.
We price the option through the Fourier--Laplace transform
\begin{equation}
\label{eq:vix_price_and_transform}
f(u;0,T)
=
\mathbb E_0^{\mathbb Q}
\left[
\exp\left\{
u\mathrm{VIX}_{T,\tau_0}^{2}
\right\}
\right].
\end{equation}

Let $R_\kappa$ denote the resolvent of $\kappa K$, satisfying
$R_\kappa+\kappa K*R_\kappa=\kappa K$.
By \autoref{lem:fractional_resolvent},
\begin{equation}
\label{eq:vix_averaged_resolvent}
R_\kappa(s)
=
\kappa s^{\alpha-1}E_{\alpha,\alpha}(-\kappa s^\alpha),
\quad
a_{\tau_0}(s)
=
\frac{1}{\kappa\tau_0}
\int_s^{s+\tau_0}R_\kappa(v)\,\mathrm{d}v.
\end{equation}
Suppose that \autoref{eq:rough_isvix_model} admits the finite-horizon solution specified in \appref{app:risvix_admissibility} on $[0,T+\tau_0]$.

\begin{theorem}
\label{thm:vix_transform}
There exists $\bar\eta_{T,\tau_0}>0$ such that, for every $u\in\mathbb C$ with $0\leq\operatorname{Re}(u)<\bar\eta_{T,\tau_0}$, the transform satisfies almost surely
\begin{equation}
\label{eq:vix_transform}
f(u;0,T)
=\exp\left\{\phi(T)+\chi_1(T)V_0+\psi_2(T)\theta_0\right\},
\end{equation}
where the coefficients $\psi_1$, $\psi_2$, $\chi_1$ and $\phi$ are the unique continuous functions on $[0,T]$ satisfying
\begin{equation}
\label{eq:vix_transform_system}
\begin{aligned}
\psi_1(s)
&=u a_{\tau_0}(s)
+\frac{\sigma_V^2}{2\kappa}\int_0^sR_\kappa(s-v)\psi_1(v)^2\,\mathrm{d}v,\\
\psi_2(s)
&=u[1-w(\tau_0)]
+\int_0^s\left[\kappa\psi_1(v)+\frac{\sigma_\theta^2}{2}\psi_2(v)^2\right]\mathrm{d}v,\\
\chi_1(s)
&=u w(\tau_0)
+\int_0^s\left[-\kappa\psi_1(v)+\frac{\sigma_V^2}{2}\psi_1(v)^2\right]\mathrm{d}v,\\
\phi(s)
&=\lambda^y\int_0^s\left[\varrho(\psi_1(v))-1-\mu_y\psi_1(v)\right]\mathrm{d}v.
\end{aligned}
\end{equation}
The solution satisfies $\sup_{0\leq s\leq T}\operatorname{Re}\psi_1(s)<\mu_y^{-1}$, so the jump transform $\varrho$ is well defined.
\end{theorem}

\noindent\textbf{Remark}.
For each transform argument $u$, computing the coefficients requires solving a nonlinear Volterra equation for $\psi_1$ and a scalar Riccati ODE for $\psi_2$; $\chi_1$ and $\phi$ are then obtained by integration. The variance-window averaging enters through $a_{\tau_0}$, while fractional memory enters through both $a_{\tau_0}$ and $R_\kappa$. Formally setting $\alpha=1$ gives $a_{\tau_0}(s)=w(\tau_0)e^{-\kappa s}$ and $\chi_1=\psi_1$, reducing \autoref{eq:vix_transform_system} to the corresponding Markovian Riccati system.

To recover option prices from \autoref{eq:vix_transform}, we view the call payoff as a function of squared VIX,
$g(x)=(\sqrt{x}-K)^+$ for $x\geq0$.
Its Laplace transform is
$\int_0^\infty e^{-zx}g(x)\,\mathrm{d}x
=
\sqrt{\pi}\operatorname{erfc}(K\sqrt z)/(2z^{3/2})$
for $\operatorname{Re}(z)>0$.
\autoref{thm:vix_transform} guarantees that the VIX transform is finite on every vertical contour
$u=\eta+i\xi$ with $\eta\in(0,\bar\eta_{T,\tau_0})$.
Hence any such $\eta$ can be used as the damping parameter in the inversion below; the resulting option value does not depend on its particular choice.

\begin{corollary}
\label{thm:vix_call_price}
Assume a constant risk-free rate $r$ and let
$\eta\in(0,\bar\eta_{T,\tau_0})$.
Then the time-zero price of a European VIX call with expiry $T$ and strike $K$ is
\begin{equation}
\label{eq:vix_call_price}
\begin{aligned}
c(0,T;K)
&=
e^{-rT}
\mathbb E_0^{\mathbb Q}
\left[
(\mathrm{VIX}_{T,\tau_0}-K)^+
\right]\\
&=
\frac{e^{-rT}}{2\sqrt{\pi}}
\int_0^\infty
\operatorname{Re}
\left[
\frac{
\operatorname{erfc}\!\left(
K\sqrt{\eta+i\xi}
\right)
}{
(\eta+i\xi)^{3/2}
}
f(\eta+i\xi;0,T)
\right]\mathrm{d}\xi.
\end{aligned}
\end{equation}
Here
$\operatorname{erfc}(z)
=
1-\frac{2}{\sqrt{\pi}}\int_0^z e^{-v^2}\,\mathrm{d}v$,
and the square root and complex power are taken on the principal branch.
\end{corollary}

Thus, VIX option pricing under RISVIX reduces to solving one nonlinear Volterra equation for each transform argument and evaluating a one-dimensional numerical integral.

\subsection{Calibration Strategy}
\label{subsec:calibration-strategy}

The parameter separation leads naturally to a two-step procedure, following the estimation--calibration structure of \citet{LZZ2019}.
The first step estimates the common term-structure parameters and date-specific initial states from VIX indices, while the second step calibrates the remaining distributional parameters to VIX option prices.

\paragraph{Step 1: Estimation from the VIX term structure}
Let $\mathcal T$ denote the set of calibration dates and let $\tau_1,\ldots,\tau_J$ denote the observed VIX horizons.
At each calendar date $t_0\in\mathcal T$, we treat $t_0$ as the valuation origin of a local model and apply the time-zero relation in
\autoref{thm:vix_term_structure},
$(\mathrm{VIX}^{\mathrm{model}}_{t_0,\tau_j})^2
=
V_{t_0}w(\tau_j)+\theta_{t_0}[1-w(\tau_j)]$.
Here $(V_{t_0},\theta_{t_0})$ are the date-specific initial states of the local model; the subscript $t_0$ indexes the calendar valuation date.
We estimate the common parameters $(\alpha,\kappa)$ jointly with these initial states by
\begin{equation}
\label{eq:term-structure-calibration}
\left(
\widehat\alpha,\widehat\kappa,
\{\widehat V_{t_0},\widehat\theta_{t_0}\}_{t_0\in\mathcal T}
\right)
=
\underset{\substack{\alpha\in(1/2,1],\,\kappa>0\\
V_{t_0}\geq0,\,\theta_{t_0}\geq0,\ t_0\in\mathcal T}}
{\arg\min}
\sum_{t_0\in\mathcal T}\sum_{j=1}^{J}
\left[
\mathrm{VIX}^{\mathrm{mkt}}_{t_0,\tau_j}
-
\mathrm{VIX}^{\mathrm{model}}_{t_0,\tau_j}
(\alpha,\kappa,V_{t_0},\theta_{t_0})
\right]^2.
\end{equation}
For any fixed $(\alpha,\kappa)$, the date-specific initial-state estimates are separable across dates and are profiled out subject to nonnegativity.
Our LZZ benchmark applies the same procedure while imposing $\alpha=1$.

\paragraph{Step 2: Calibration to VIX option prices}
Conditional on the Step 1 estimates, we calibrate
$\boldsymbol\zeta
=
(\sigma_V,\sigma_\theta,\lambda^y,\mu_y)^\top$
using VIX call prices.
Let $\mathcal T_{\mathrm{train}}$ denote the set of option-calibration dates.
For each $t_0\in\mathcal T_{\mathrm{train}}$, let $\mathcal M_{t_0}$ denote the observed times to expiration and let $\mathcal K_{t_0,T}$ denote the strikes available for $T\in\mathcal M_{t_0}$.
The local model at $t_0$ is initialized with $(\widehat V_{t_0},\widehat\theta_{t_0})$.
The option maturity $T$ is measured from $t_0$, and the VIX variance horizon $\tau_0$ is fixed at 30 days.
The second-step estimator is
\begin{equation}
\label{eq:option-calibration-sse}
\widehat{\boldsymbol\zeta}
=
\underset{\boldsymbol\zeta\in\mathcal P}{\arg\min}
\sum_{t_0\in\mathcal T_{\mathrm{train}}}
\sum_{T\in\mathcal M_{t_0}}
\sum_{K\in\mathcal K_{t_0,T}}
\left[
C_{t_0}^{\mathrm{mkt}}(T,K)
-
C_{t_0}^{\mathrm{model}}
\left(
T,K;
\boldsymbol\zeta,
\widehat\alpha,\widehat\kappa,
\widehat V_{t_0},\widehat\theta_{t_0}
\right)
\right]^2.
\end{equation}
Here $C_{t_0}^{\mathrm{mkt}}(T,K)$ is the bid--ask midpoint, and
$C_{t_0}^{\mathrm{model}}(T,K;\cdot)$ is the price
$c(0,T;K)$ in \autoref{eq:vix_call_price}, evaluated using the Step 1
estimates and the candidate value of $\boldsymbol\zeta$.
The parameters in $\boldsymbol\zeta$ are common across all valuation dates,
expirations and strikes.

For each valuation date $t_0$, we re-index that date as local time zero.
Under this re-indexing, the relevant local model interval for an option with time to expiration $T$ is $[0,T+\tau_0]$, corresponding to the calendar interval $[t_0,t_0+T+\tau_0]$.
Such re-indexing is standard in rough-volatility pricing; depending on the model, conditioning at the new origin is encoded through an initial forward curve or another model-specific initial input \citep{ER2018,HJT2020}.
In our empirical implementation, each $t_0$ defines a local RISVIX valuation problem initialized by the date-specific initial states $(\widehat V_{t_0},\widehat\theta_{t_0})$ inferred from the contemporaneous VIX term structure.

\section{Empirical Study}
\label{sec4}

\subsection{Data}
\label{subsec:data}

We obtain daily VIX term-structure observations from CBOE and VIX option quotes from OptionMetrics.
The term structure consists of the closing values of VIX9D, VIX, VIX3M, VIX6M and VIX1Y, corresponding to volatility horizons of nine days, 30 days, three months, six months and one year, respectively.
The sample runs from January 2, 2018 to June 30, 2025.
Restricting attention to dates on which all five indices are observed gives a balanced panel of 1,883 trading days.

The VIX option data also run from January 2, 2018 to June 30, 2025 and contain trading dates, expiration dates, strikes, option types, closing bid and ask quotes, and trading volume.
We use the bid--ask midpoint as the observed option price, with option prices and strikes expressed in VIX index points.
We retain call options with 30--126 calendar days to expiration and spot moneyness $K/\mathrm{VIX}_t\in [0.7,1.3]$, where $K$ denotes the strike and $\mathrm{VIX}_t$ is the contemporaneous closing value of the 30-day VIX index.
We further require bid and ask quotes of at least 0.025 points, positive trading volume and open interest, a midquote of at least 0.50 points, a nonnegative bid--ask spread no greater than 10\% of the midquote.
These filters leave 61,994 quote--day observations: 47,451 in the training period from January 2018 to December 2023 and 14,543 in the test period from January 2024 to June 2025.

\begin{table}[!htbp]
\centering
\caption{\textbf{Summary statistics of VIX indices and options.}
\footnotesize{Panel A reports daily closing index levels.
Panel B reports  VIX call-option midquotes, grouped by calendar days to expiration.}}
\label{tab:vix-data-summary}
\label{tab:vix-term-structure-summary}
\label{tab:vix-option-summary}
\small
\setlength{\tabcolsep}{4pt}
\renewcommand{\arraystretch}{1.35}
\begin{tabular*}{\linewidth}{@{\extracolsep{\fill}}lrrrrrrrr@{}}
\toprule
\multicolumn{9}{l}{\textbf{Panel A: VIX term structure}} \\
\midrule
\textbf{Index}
& \textbf{Mean}
& \textbf{Std.}
& \textbf{Min.}
& \textbf{Q1}
& \textbf{Median}
& \textbf{Q3}
& \textbf{Max.}
& \textbf{Count} \\
\midrule
VIX9D & 19.29 & 9.50 & 7.70  & 13.40 & 17.04 & 22.20 & 106.66 & 1,883 \\
VIX   & 19.95 & 7.65 & 9.15  & 14.74 & 18.20 & 22.88 & 82.69  & 1,883 \\
VIX3M & 21.77 & 6.72 & 12.06 & 16.60 & 20.40 & 25.31 & 72.98  & 1,883 \\
VIX6M & 22.97 & 5.94 & 14.11 & 18.08 & 21.81 & 26.88 & 61.11  & 1,883 \\
VIX1Y & 23.55 & 4.92 & 15.62 & 19.40 & 22.37 & 27.82 & 45.86  & 1,883 \\
\midrule
\multicolumn{9}{l}{\textbf{Panel B: VIX call options}} \\
\midrule
\multicolumn{2}{l}{\textbf{Maturity}}
& \textbf{Mean}
& \textbf{Std.}
& \textbf{Min.}
& \textbf{Max.}
& \textbf{Skew.}
& \textbf{Kurt.}
& \textbf{Count} \\
\midrule
\multicolumn{2}{l}{All}
& 4.28 & 2.14 & 0.53 & 15.90 & 0.89 & 3.57 & 61,994 \\
\multicolumn{2}{l}{30--59 days}
& 3.74 & 2.06 & 0.53 & 15.90 & 1.10 & 4.24 & 21,232 \\
\multicolumn{2}{l}{60--126 days}
& 4.56 & 2.12 & 0.53 & 14.60 & 0.83 & 3.41 & 40,762 \\
\bottomrule
\end{tabular*}
\end{table}

\autoref{tab:vix-data-summary} reports summary statistics for the two datasets.
Panel A shows an upward-sloping average term structure, with mean index levels increasing from 19.29 for VIX9D to 23.55 for VIX1Y, while the corresponding standard deviations decrease from 9.50 to 4.92.
Panel B reports an average option midquote of 4.28 points.
Longer-dated calls have higher average prices, whereas shorter-dated calls exhibit greater positive skewness and kurtosis.

\begin{figure}[!htbp]
    \centering
    \begin{minipage}[b]{0.53\linewidth}
        \centering
        \includegraphics[width=\linewidth]{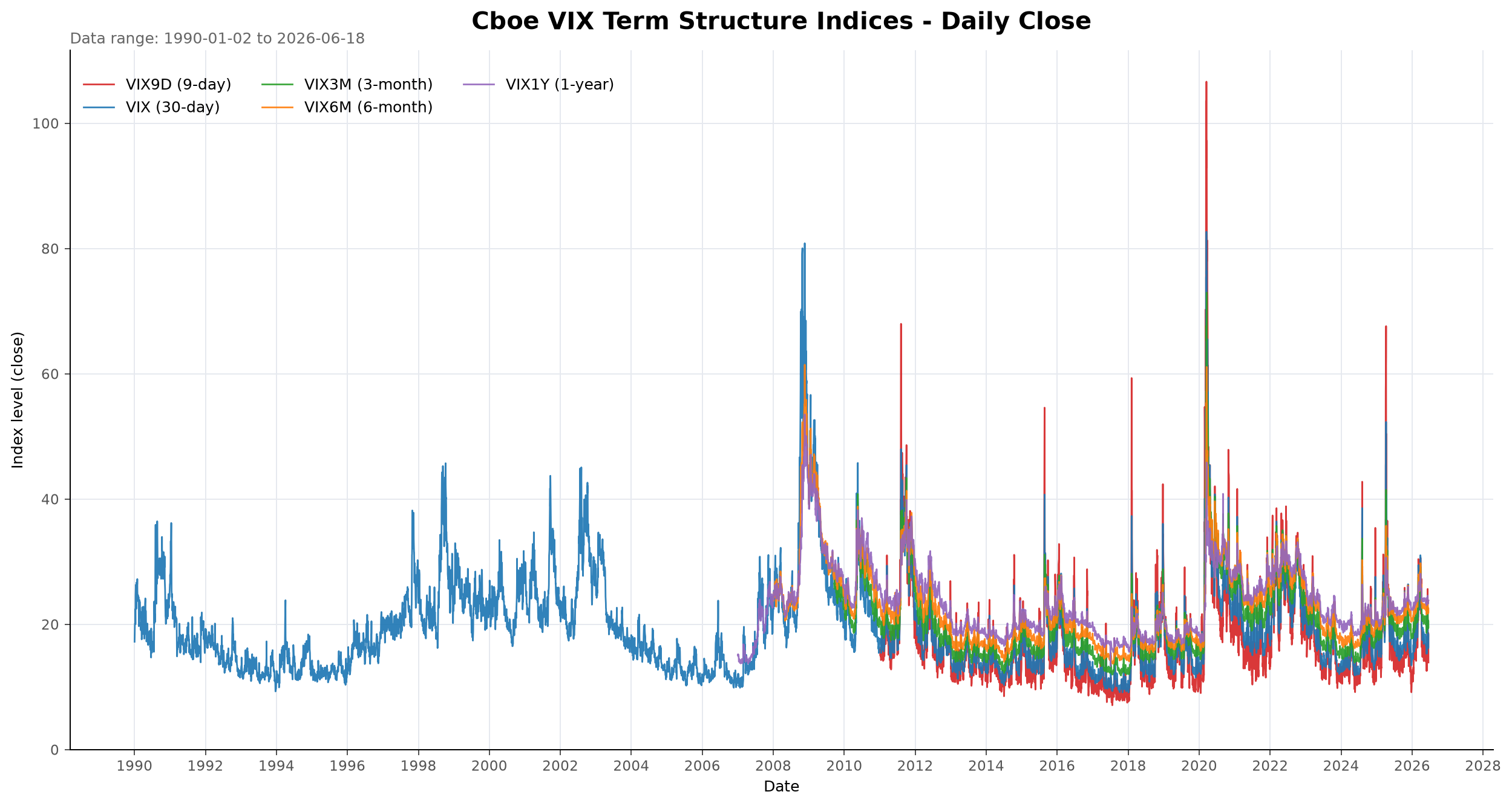}
        \par\smallskip
        {\small (a) Historical index dynamics}
    \end{minipage}\hfill
    \begin{minipage}[b]{0.42\linewidth}
        \centering
        \includegraphics[width=\linewidth]{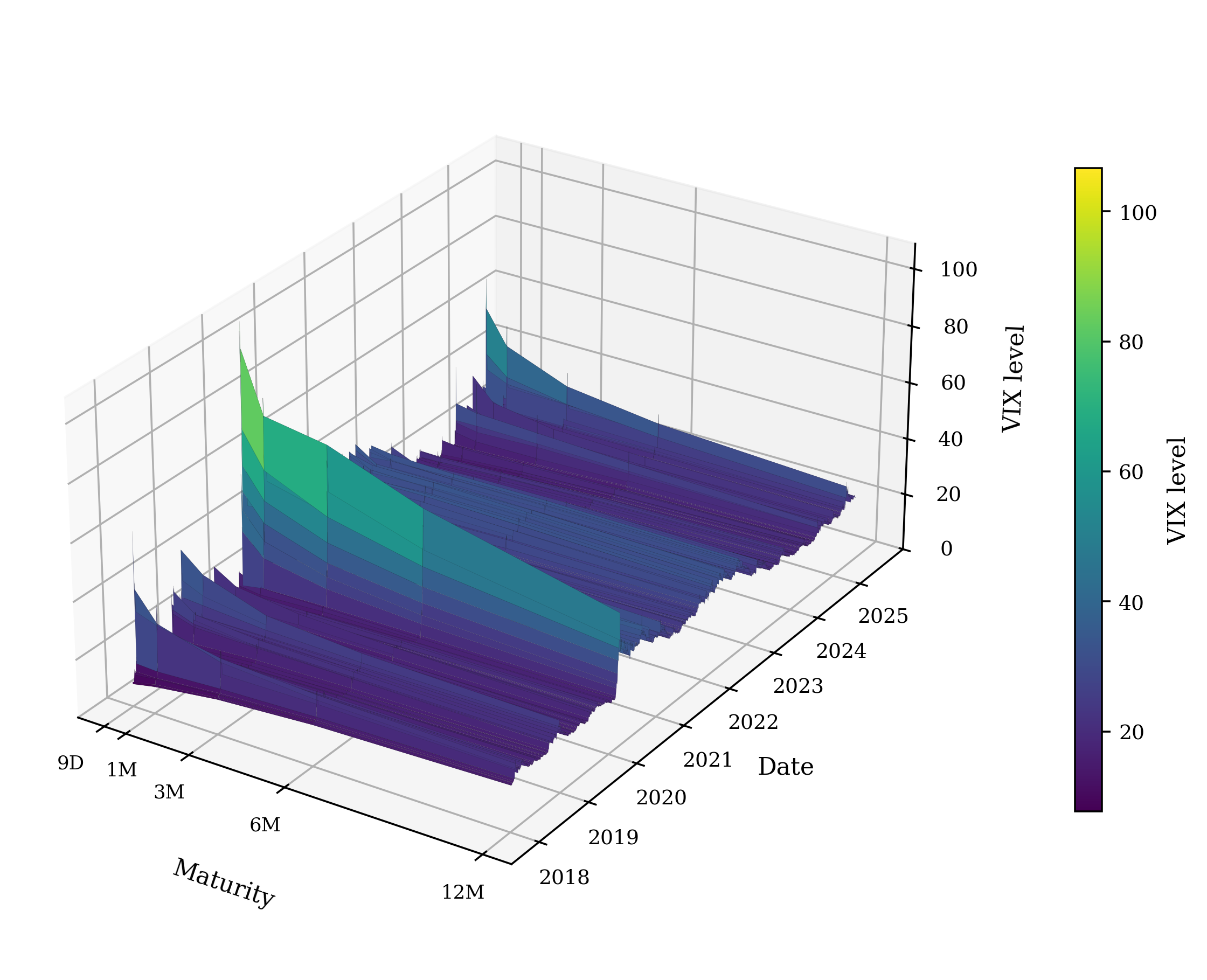}
        \par\smallskip
        {\small (b) VIX term structure}
    \end{minipage}
    \caption{\textbf{Historical dynamics and term structure of VIX indices.}
    \footnotesize{Panel (a) plots the available daily closing values of VIX9D, VIX, VIX3M, VIX6M and VIX1Y on a common scale.
    Panel (b) displays the balanced term-structure panel used in the empirical analysis.}}
    \label{fig:vix-term-structure-data}
    \label{fig:market-term-structure-surface}
\end{figure}

\autoref{fig:vix-term-structure-data} illustrates the historical evolution and term-structure variation of the VIX indices. 
Panel (a) shows substantial co-movement across horizons, with shorter-horizon indices exhibiting sharper fluctuations.
Panel (b) shows that the term structure is typically upward sloping during low-volatility periods but can invert during volatility spikes, most notably in early 2020. 
Thus, the data exhibit substantial variation in both the level and shape of the VIX term structure.

\subsection{Term-Structure Calibration}
\label{sec:ts-calibration}

We compare RISVIX with the benchmark of \citet{LZZ2019}, denoted LZZ ($\alpha=1$), using the VIX-level least-squares criterion in \autoref{sec3}.
The full-sample comparison also includes the rough Heston term-structure model of \citet{YeFanKwok2026}, denoted YFK.
The full calibration sample in this subsection covers January 2018--June 2025 and contains 1,883 trading days and 9,415 horizon--date observations.
We also examine a separate calibration of RISVIX and LZZ over January--June 2020.
For the subsequent option pricing, the first-step inputs are estimated separately using January 2018--December 2023, and January 2024--June 2025 is reserved as the test period.

For RISVIX, the common parameters $(\alpha,\kappa)$ and date-specific nonnegative initial states $(V_t,\theta_t)$ are updated alternately.
Given $(\alpha,\kappa)$, we fit $(V_t,\theta_t)$ separately for each date; holding those states fixed, we then update $(\alpha,\kappa)$ using all calibration dates.
The states are re-estimated at the updated parameters, and this cycle is repeated until the stopping tolerance or iteration limit is reached.
LZZ imposes $\alpha=1$ and uses the same VIX-level least-squares criterion.
\subsubsection{Full-Sample Fit}
\label{subsubsec:ts-full-sample}

\autoref{tab:ts-rough-lzz-full-sample} reports the parameter estimates and calibration errors for January 2018--June 2025.
We recalibrate YFK \citep{YeFanKwok2026} to the same VIX term-structure observations using a 20-factor Markovian lift of the fractional volatility kernel.

In addition to VIX levels, we assess term-structure slope and curvature using divided differences.
Let $y_{t,j}$ denote the VIX index level on date $t$ at horizon $\tau_j$, where $\tau_1<\cdots<\tau_5$ are expressed in years.
We approximate slope and curvature by
\begin{equation}
\begin{aligned}
s_{t,j}
&=\frac{y_{t,j+1}-y_{t,j}}{\tau_{j+1}-\tau_j},
&& j=1,\ldots,4,\\
c_{t,j}
&=\frac{2\bigl(s_{t,j+1}-s_{t,j}\bigr)}{\tau_{j+2}-\tau_j},
&& j=1,\ldots,3.
\end{aligned}
\end{equation}
These definitions account for unequal horizon spacings and are applied to both observed and model-implied VIX levels.

For each measure, MSE is the average squared difference between observed and model-implied values over the relevant dates and horizons.
Each model is evaluated on 9,415 level errors, 7,532 slope errors and 5,649 curvature errors.
The percentage MSE reduction relative to benchmark $B\in\{\mathrm{LZZ},\mathrm{YFK}\}$ is $100(1-\mathrm{MSE}_{\mathrm{RISVIX}}/\mathrm{MSE}_B)$; positive values favour RISVIX.

\begin{table}[!htbp]
\centering
\caption{\textbf{Full-sample term-structure calibration.}
\footnotesize{Panel A reports parameter estimates for January 2018--June 2025 (1,883 trading days).
LZZ imposes $\alpha=1$, and $\kappa$ corresponds to the parameter denoted by $\lambda$ in YFK.
Panel B reports MSE for the three models on the same observations and the percentage reductions achieved by RISVIX relative to LZZ and YFK.
Positive reductions favour RISVIX.}}
\label{tab:ts-rough-lzz-full-sample}
\small
\setlength{\tabcolsep}{3pt}
\renewcommand{\arraystretch}{1.35}

\begin{tabular*}{\linewidth}{@{\extracolsep{\fill}}lrrr@{}}
\toprule
\multicolumn{4}{@{}l}{\textbf{Panel A: Parameter estimates}} \\
\midrule
\textbf{Parameter}
& \textbf{RISVIX}
& \textbf{LZZ}
& \textbf{YFK} \\
\midrule
$\widehat\alpha$
& 0.8942 & 1.0000 & 0.6997 \\
$\widehat\kappa$
& 6.8509 & 6.0322 & 1.3368 \\
\bottomrule
\end{tabular*}

\medskip

\begin{tabular*}{\linewidth}{@{\extracolsep{\fill}}lrrrrr@{}}
\toprule
\multicolumn{6}{@{}l}{\textbf{Panel B: Calibration errors}} \\
\midrule
& & & &
\multicolumn{2}{c@{}}{\textbf{MSE reduction (\%)}} \\
\cmidrule(l){5-6}
\textbf{Metric}
& \textbf{RISVIX}
& \textbf{LZZ}
& \textbf{YFK}
& \textbf{vs.\ LZZ}
& \textbf{vs.\ YFK} \\
\midrule
VIX level
& 0.5639 & 0.6295 & 4.8576
& 10.42 & 88.39 \\
Slope
& 225.8782 & 272.8391 & 536.0767
& 17.21 & 57.86 \\
Curvature
& 27627.1098 & 31822.2198 & 38021.9485
& 13.18 & 27.34 \\
\bottomrule
\end{tabular*}
\end{table}

RISVIX yields $\widehat\alpha=0.8942$ ($\widehat H=0.3942$), compared with $\widehat\alpha=0.6997$ ($\widehat H=0.1997$) for YFK, and achieves the lowest MSE across all three measures.
Relative to LZZ, RISVIX reduces level MSE by 10.42\%, with larger reductions of 17.21\% and 13.18\% for slope and curvature, respectively.
The corresponding reductions relative to YFK are 88.39\%, 57.86\% and 27.34\%.

\subsubsection{State-Dependent Fit}
\label{subsubsec:ts-state-dependence}

\begin{figure}[!htbp]
\centering
\includegraphics[width=0.97\linewidth]{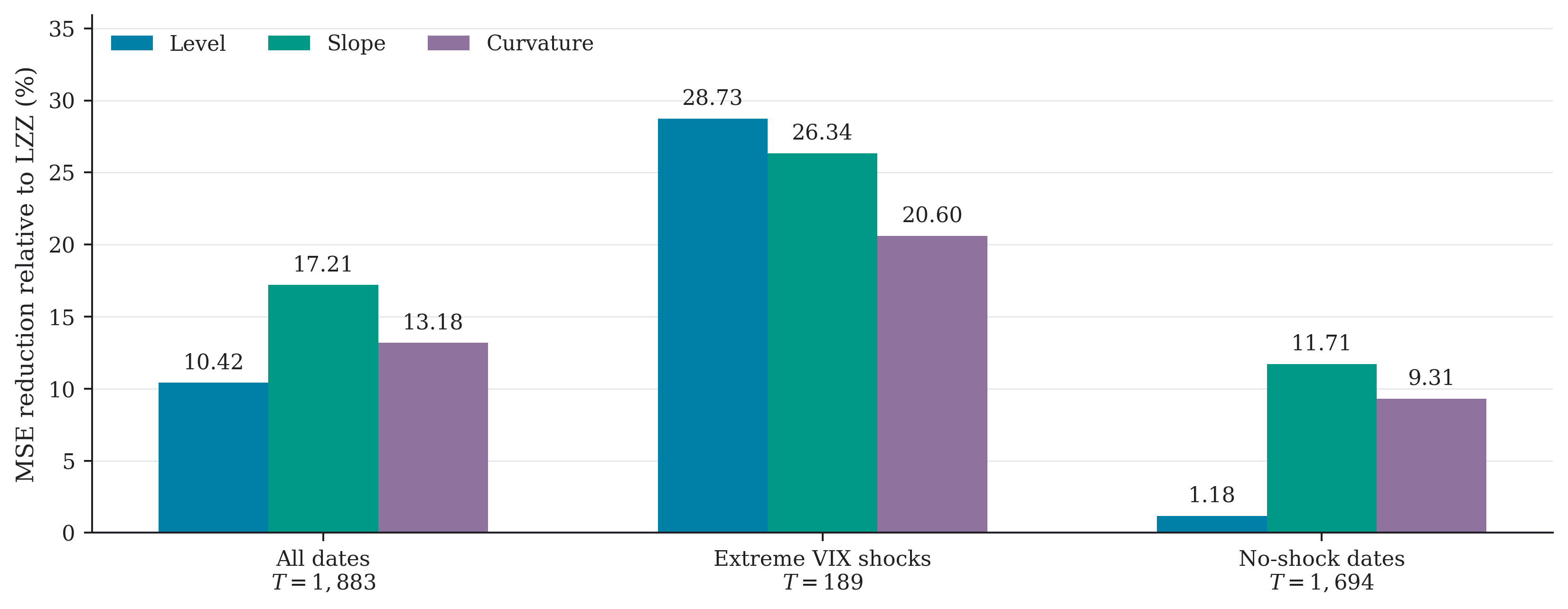}
\caption{\textbf{Term-structure fit by VIX shocks.}
\footnotesize{Bars report percentage MSE reductions for RISVIX relative to LZZ for each diagnostic and group. The sample runs from January 2, 2018 to June 30, 2025. }}
\label{fig:ts-rough-lzz-shock-comparison}
\end{figure}

The relative gains of RISVIX vary substantially across market conditions.
We split the full calibration sample into dates with large VIX increases and the remaining dates, keeping the full-sample model parameters fixed.
A shock date is defined as a daily VIX increase at or above the 90th percentile of the calibration-sample changes.
\autoref{fig:ts-rough-lzz-shock-comparison} reports percentage MSE reductions for levels, slopes and curvature in the full sample and in the two groups.
On shock dates, the respective reductions are 28.73\%, 26.34\% and 20.60\%, compared with 1.18\%, 11.71\% and 9.31\% on the remaining dates.
RISVIX therefore improves all three diagnostics in both groups, with larger relative improvements when VIX rises sharply.

We next examine the initial COVID-19 period from January 2 to June 30, 2020, which contains 125 trading days and 625 horizon--date observations.
Both models are separately estimated using this period.

\begin{table}[!htbp]
\centering
\caption{\textbf{Term-structure calibration during the COVID-19 period.}
\footnotesize{Panel A reports parameter values for January 2--June 30, 2020 (125 trading days); LZZ imposes $\alpha=1$.
Panel B reports MSE on identical observations for the two models and its percentage reduction relative to LZZ.
Positive reductions favour RISVIX.}}
\label{tab:ts-rough-lzz-covid}
\small
\setlength{\tabcolsep}{3pt}
\renewcommand{\arraystretch}{1.35}
\begin{tabular*}{\linewidth}{@{\extracolsep{\fill}}lrr@{\hspace{0.045\linewidth}}lrrr@{}}
\toprule
\multicolumn{3}{@{}l}{\textbf{Panel A: Parameter estimates}} & \multicolumn{4}{l@{}}{\textbf{Panel B: Calibration errors}} \\
\cmidrule(r){1-3}\cmidrule(l){4-7}
\textbf{Parameter} & \textbf{RISVIX} & \textbf{LZZ} & \textbf{Metric} & \textbf{RISVIX} & \textbf{LZZ} & \textbf{MSE reduction (\%)} \\
\cmidrule(r){1-3}\cmidrule(l){4-7}
$\widehat\alpha$ & 0.7051 & 1.0000 & VIX level & 2.4808 & 3.0105 & 17.60 \\
$\widehat\kappa$ & 8.5273 & 12.4076 & Slope & 529.0738 & 898.8543 & 41.14 \\
 & &  & Curvature & 70100.9904 & 131260.4861 & 46.59 \\
\bottomrule
\end{tabular*}
\end{table}

\autoref{tab:ts-rough-lzz-covid} shows a larger departure from the Markovian restriction during the COVID-19 period, with $\widehat\alpha=0.7051$ ($\widehat H=0.2051$).
The level MSE falls from 3.0105 for LZZ to 2.4808 for RISVIX, a reduction of 17.60\%.
The corresponding reductions in slope and curvature MSE are 41.14\% and 46.59\%.
Relative to the full calibration sample, the improvement is particularly pronounced for the shape diagnostics.

\subsubsection{Fit across Horizons}
\label{subsubsec:ts-horizons}

\begin{figure}[!htbp]
\centering
\includegraphics[width=\linewidth]{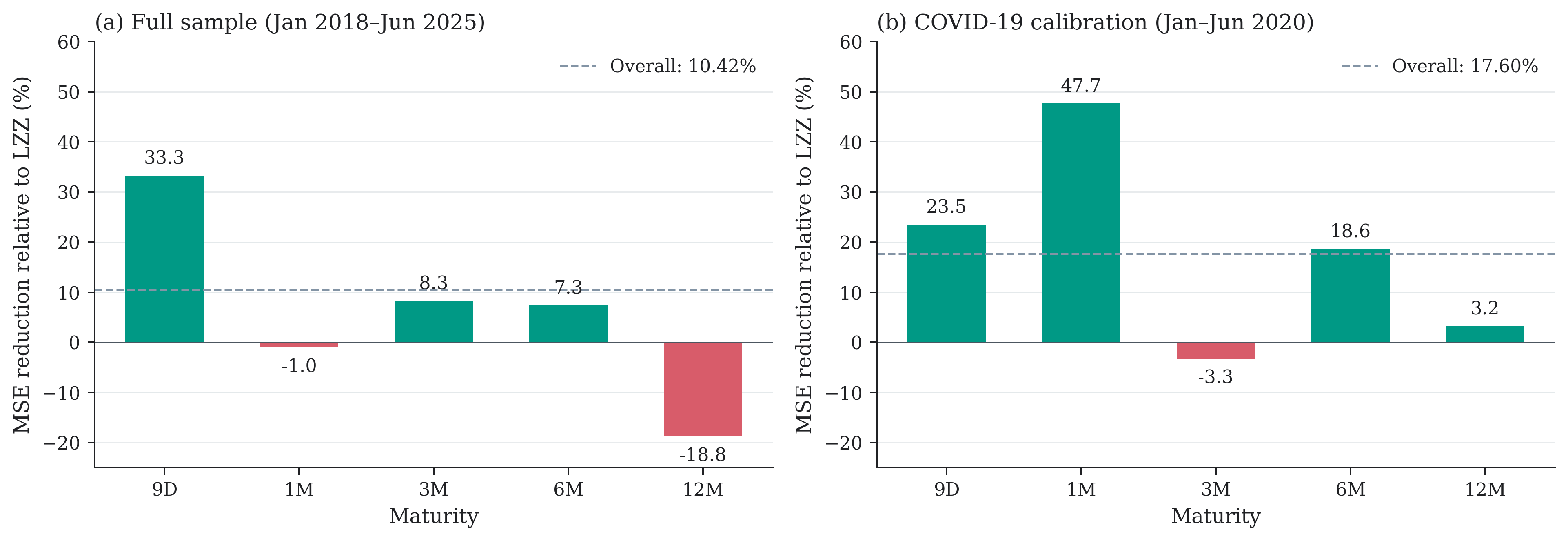}
\caption{\textbf{Term-structure fit across horizons.}
\footnotesize{Bars report horizon-specific percentage reductions in level MSE for RISVIX relative to LZZ. Panel (a) uses the full January 2018--June 2025 sample; Panel (b) uses models separately calibrated over January--June 2020. Each panel has one bar per horizon. Green bars indicate positive reductions and red bars indicate negative reductions. Dashed lines show the corresponding aggregate level-MSE reductions. The panels share the same vertical scale.}}
\label{fig:ts-rough-lzz-maturity-gains}
\end{figure}

\autoref{fig:ts-rough-lzz-maturity-gains} decomposes the percentage reduction in level MSE across VIX horizons.
Panel (a) uses the full January 2018--June 2025 sample.
The largest reduction is 33.30\% at nine days, followed by 8.26\% at three months and 7.34\% at six months.
At 30 days and one year, MSE is higher than LZZ, with reductions of $-1.05\%$ and $-18.83\%$, respectively.

Panel (b) uses the separate COVID-19 calibrations and displays one bar per horizon.
The largest reduction, 47.69\%, occurs at the 30-day horizon, followed by 23.47\% at nine days and 18.64\% at six months.
The one-year MSE declines by 3.20\%, while the three-month MSE increases by 3.33\%.
The dashed lines mark the overall level-MSE reductions of 10.42\% and 17.60\% for the two calibration samples.

\subsection{VIX Option Pricing}
\label{subsec:option-pricing}
\label{subsec:option-calibration}

We next evaluate whether the term-structure flexibility of RISVIX is associated with more accurate out-of-sample VIX option prices.
The training period is January 2018--December 2023, and the test period is January 2024--June 2025.
The global parameter sets reported below are common to all dates, expirations and strikes and are held fixed throughout the test period.
For each test date $t_0$, the local initial states $(V_{t_0},\theta_{t_0})$ are inferred from the contemporaneous VIX term structure with $(\alpha,\kappa)$ fixed at their training-sample values.

\subsubsection{Parameter Estimates}
\label{subsubsec:option-design}

The first-step inputs comprise the common term-structure parameters $(\alpha,\kappa)$ and the date-specific nonnegative states $(V_{t_0},\theta_{t_0})$.
RISVIX obtains these inputs through the alternating updates described above.
Conditional on these inputs, the option-pricing formula uses the four distributional parameters $(\sigma_V,\sigma_\theta,\lambda^y,\mu_y)$.
\autoref{tab:option-parameters} reports the global parameter values used to generate the option prices in this subsection.
RISVIX uses $\widehat\alpha=0.8937$ ($\widehat H=0.3937$) and $\widehat\kappa=7.0357$, while LZZ imposes $\alpha=1$ and uses $\widehat\kappa=6.2450$.
Relative to LZZ, RISVIX has higher $\widehat\sigma_V$ and $\widehat\sigma_\theta$, a higher jump intensity $\widehat\lambda^y$, and a smaller mean jump size $\widehat\mu_y$.

\begin{table}[!htbp]
\centering
\caption{\textbf{Option-pricing parameters.}
\footnotesize{The table reports the global parameter values used for the option-pricing results. The term-structure inputs $(\alpha,H,\kappa)$ are based on the January 2018--December 2023 term structure, with $H=\alpha-1/2$. All global parameters are held fixed over the test period.}}
\label{tab:option-parameters}
\small
\setlength{\tabcolsep}{4pt}
\renewcommand{\arraystretch}{1.4}
\begin{tabular*}{\linewidth}{@{\extracolsep{\fill}}lrrrrrrr@{}}
\toprule
\textbf{Model} & $\widehat\alpha$ & $\widehat H$ & $\widehat\kappa$ & $\widehat\sigma_V$ & $\widehat\sigma_\theta$ & $\widehat\lambda^y$ & $\widehat\mu_y$ \\
\midrule
RISVIX & 0.8937 & 0.3937 & 7.0357 & 1.6298 & 0.2889 & 0.0178 & 2.7920 \\
LZZ & 1.0000 & 0.5000 & 6.2450 & 1.2422 & 0.1994 & 0.0161 & 3.5847 \\
\bottomrule
\end{tabular*}
\end{table}

\subsubsection{Pricing Performance}
\label{subsubsec:option-performance}

\autoref{tab:option-pricing-errors} reports mean squared error (MSE) and mean absolute error (MAE) for the complete option sample and for maturity and moneyness groups.
Moneyness is measured as $K/\mathrm{VIX}_t$ and grouped as $<0.9$, $\in[0.9,1.1]$, and $>1.1$.
The two models are evaluated on the same quotes in every comparison.

\begin{table}[!htbp]
\centering
\caption{\textbf{VIX option-pricing errors.}
\footnotesize{Panel A uses 47,451 training-period quotes; Panel B uses 14,543 test-period quotes. MSE and MAE are computed from option prices in VIX index points. Red.\ is the percentage error reduction relative to LZZ; positive values favour RISVIX. The lower error in each model pair is in bold.}}
\label{tab:option-pricing-errors}
\small
\setlength{\tabcolsep}{3.5pt}
\renewcommand{\arraystretch}{1.35}
\begin{tabular*}{\linewidth}{@{\extracolsep{\fill}}lrrrrrr@{}}
\toprule
& \multicolumn{3}{c}{\textbf{MSE}} & \multicolumn{3}{c}{\textbf{MAE}} \\
\cmidrule(lr){2-4}\cmidrule(l){5-7}
\textbf{Group} & \textbf{RISVIX} & \textbf{LZZ} & \textbf{Red. (\%)} & \textbf{RISVIX} & \textbf{LZZ} & \textbf{Red. (\%)} \\
\midrule
\multicolumn{7}{l}{\textbf{Panel A: In-sample}} \\
All & \textbf{0.5015} & 0.5548 & 9.61 & \textbf{0.5528} & 0.5815 & 4.93 \\
30--59 days & \textbf{0.6090} & 0.6682 & 8.86 & \textbf{0.6271} & 0.6555 & 4.33 \\
60--126 days & \textbf{0.4455} & 0.4957 & 10.13 & \textbf{0.5141} & 0.5429 & 5.31 \\
Moneyness $<0.9$ & \textbf{0.6409} & 0.6473 & 1.00 & \textbf{0.6009} & 0.6065 & 0.94 \\
Moneyness $\in[0.9,1.1]$ & \textbf{0.4273} & 0.4981 & 14.21 & \textbf{0.5088} & 0.5500 & 7.48 \\
Moneyness $>1.1$ & \textbf{0.4563} & 0.5335 & 14.47 & \textbf{0.5587} & 0.5946 & 6.04 \\
\midrule
\multicolumn{7}{l}{\textbf{Panel B: Out-of-sample}} \\
All & \textbf{0.6293} & 0.7238 & 13.06 & \textbf{0.7061} & 0.7537 & 6.33 \\
30--59 days & \textbf{0.7204} & 0.8485 & 15.10 & \textbf{0.7643} & 0.8203 & 6.83 \\
60--126 days & \textbf{0.5820} & 0.6590 & 11.69 & \textbf{0.6758} & 0.7192 & 6.03 \\
Moneyness $<0.9$ & \textbf{0.4536} & 0.5945 & 23.71 & \textbf{0.5555} & 0.6324 & 12.16 \\
Moneyness $\in[0.9,1.1]$ & \textbf{0.7243} & 0.8299 & 12.73 & \textbf{0.7890} & 0.8458 & 6.72 \\
Moneyness $>1.1$ & \textbf{0.7071} & 0.7361 & 3.93 & \textbf{0.7709} & 0.7746 & 0.48 \\
\bottomrule
\end{tabular*}
\end{table}

In sample, RISVIX lowers overall MSE from 0.5548 to 0.5015 and MAE from 0.5815 to 0.5528, corresponding to reductions of 9.61\% and 4.93\%.
Both error measures are lower in every reported group, although the reductions are small for moneyness below $0.9$.
Out of sample, the overall MSE falls from 0.7238 to 0.6293 and MAE from 0.7537 to 0.7061, reductions of 13.06\% and 6.33\%.
The shorter-maturity group has larger test-period gains, with MSE and MAE reductions of 15.10\% and 6.83\%, compared with 11.69\% and 6.03\% at 60--126 days.
Across moneyness, RISVIX improves both measures in all three groups; the largest test-period reductions occur below $0.9$, at 23.71\% for MSE and 12.16\% for MAE.
The gains are smaller above $1.1$, where MSE and MAE decline by 3.93\% and 0.48\%.

\begin{figure}[!htbp]
\centering
\includegraphics[width=\linewidth]{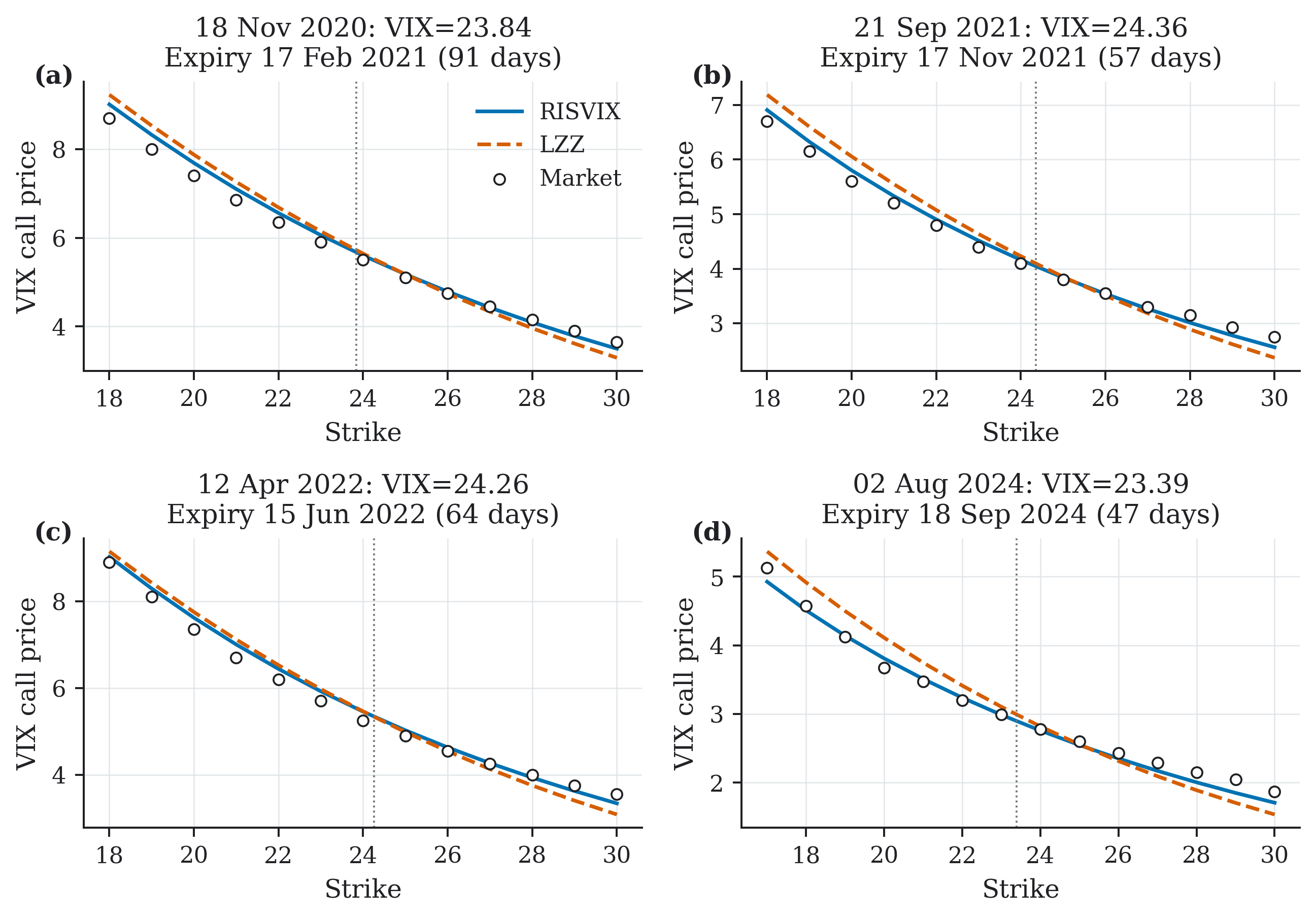}
\caption{\textbf{Cross-sectional VIX option fits.}
\footnotesize{Each panel compares market midquotes with RISVIX and LZZ prices across strikes for a fixed date and expiration. The vertical dotted line marks observed spot VIX. Panels (a)--(c) are in-sample; Panel (d) is out-of-sample. Model prices use the parameter values in \autoref{tab:option-parameters}.}}
\label{fig:option-cross-sections}
\end{figure}

\autoref{fig:option-cross-sections} provides selected price cross sections for three training-period dates and one test-period date.
Both models reproduce the decline in call prices across strikes.
In these examples, LZZ tends to lie above market prices below spot VIX and below market prices at higher strikes.
RISVIX reduces this tilt and has a lower mean squared error in each displayed cross section.
The fitted curves remain close near the centre of the strike range, with more visible differences toward its edges.
These selected examples illustrate cross-sectional shape; the complete-sample comparisons in \autoref{tab:option-pricing-errors} provide the aggregate model ranking.

\subsubsection{Pricing Performance across VIX States}
\label{subsubsec:option-states}

\begin{figure}[!htbp]
\centering
\includegraphics[width=\linewidth]{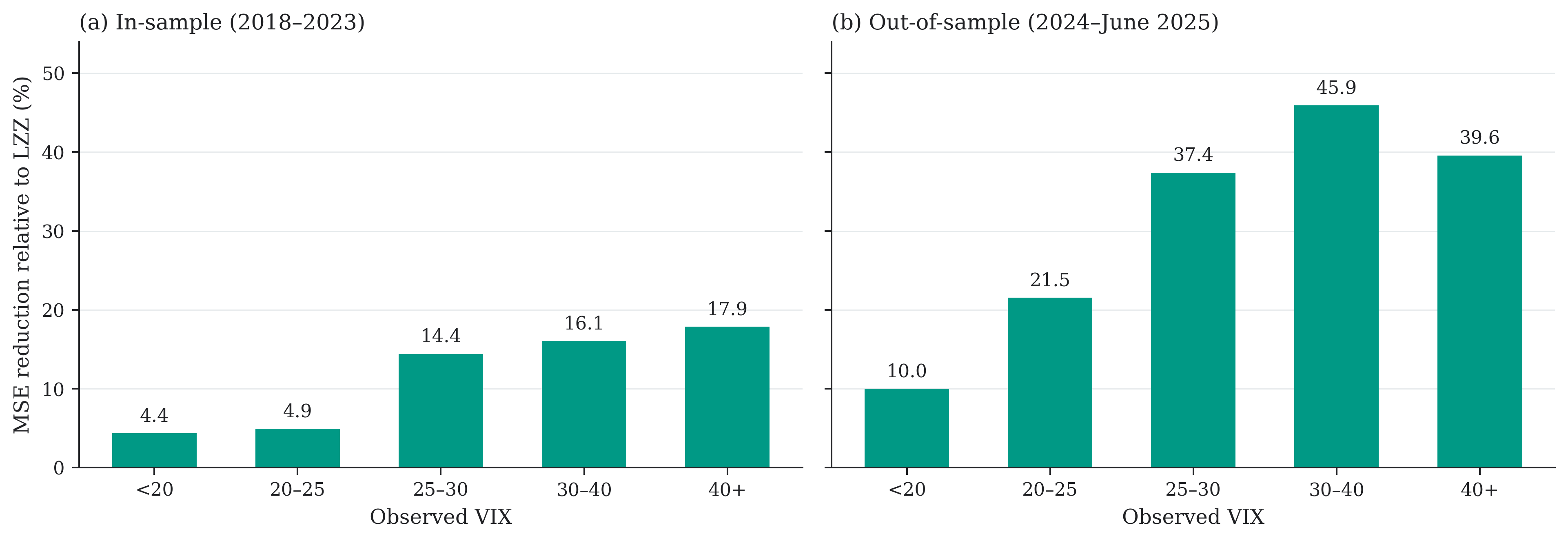}
\caption{\textbf{Option-pricing fit across VIX states.}
\footnotesize{Bars report quote-weighted percentage MSE reductions for RISVIX relative to LZZ across bins of observed VIX. Panel (a) is in-sample and Panel (b) is out-of-sample. 
Bins are left-closed and right-open, and the two panels share a vertical scale; positive values favour RISVIX.}}
\label{fig:option-vix-bins}
\end{figure}

Given the state dependence in term-structure fit, we next examine whether a similar pattern appears in VIX option prices.
\autoref{fig:option-vix-bins} reports quote-weighted MSE reductions by the observed VIX level.
Within each bin, MSE is computed by pooling all quote errors before taking the ratio between models.

In sample, the reductions are 4.4\% below VIX 20 and 4.9\% between 20 and 25, rising to 14.4\% between 25 and 30, 16.1\% between 30 and 40, and 17.9\% at VIX 40 and above.
Out of sample, RISVIX also improves every bin: MSE reductions are 10.0\%, 21.5\%, 37.4\%, 45.9\% and 39.6\%, respectively.
The largest test-period gain occurs in the 30--40 bin.

The broader split at VIX 25 supports the same pattern.
For $\mathrm{VIX}_t\geq25$, RISVIX reduces MSE by 15.67\% in sample and 40.93\% out of sample, compared with overall reductions of 9.61\% and 13.06\%.
The larger gains in elevated-volatility states are consistent with the stronger term-structure improvements during shock dates and the COVID-19 period.

\subsubsection{Implied-Volatility Fit}
\label{subsubsec:option-iv}

\begin{figure}[!htbp]
\centering
\includegraphics[width=\linewidth]{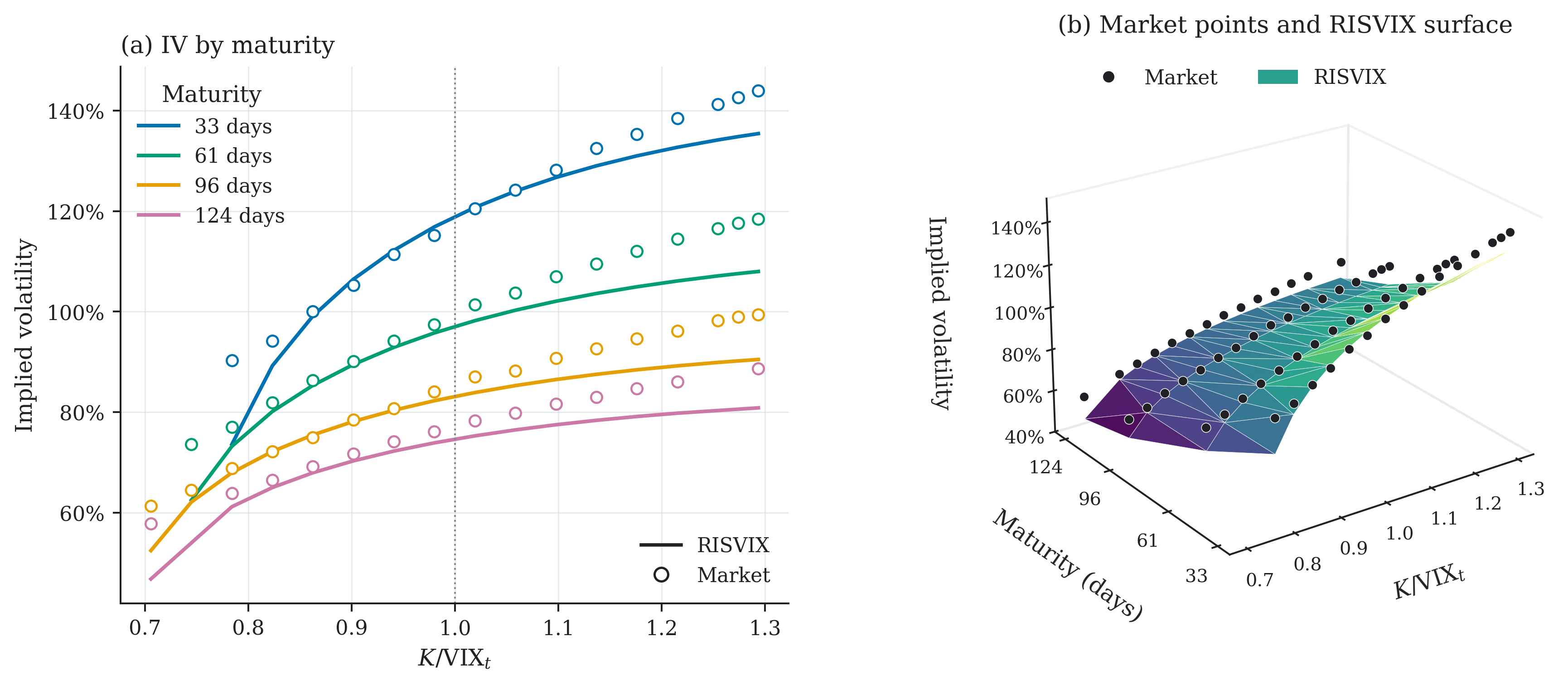}
\caption{\textbf{VIX option implied-volatility surface.}
\footnotesize{Panel (a) compares market and RISVIX implied volatilities across moneyness for four maturities on March 17, 2023. Panel (b) shows the corresponding market points and the interpolated RISVIX surface.}}
\label{fig:option-iv-surface}
\end{figure}

We complement the price-based evaluation with an implied-volatility representation for March 17, 2023.
\autoref{fig:option-iv-surface} compares market implied volatilities with those obtained from the RISVIX prices for this date.
For the Black conversion of model prices, each expiration uses the median forward inferred from its market midquotes and reported implied volatilities, with zero discounting; these conversion inputs are not observed futures quotes.

Panel (a) reports slices at 33, 61, 96 and 124 days to expiration.
Market and RISVIX implied volatilities both increase with $K/\mathrm{VIX}_t$ over the displayed ranges.
The maturity dependence is visible in the different slopes and levels of the curves.
RISVIX tracks the upward slope across all four maturities, although its implied volatilities tend to lie below the market values at higher moneyness.
The plot uses 62 quotes for which the market and both models have valid implied volatilities under the conversion convention.

Panel (b) displays the same observations as market points and a piecewise-linear RISVIX surface across moneyness and maturity.
The surface represents an interpolation of the model values at the observed contracts.
The date is selected to illustrate four well-populated, upward-sloping maturity slices.

\subsubsection{Comparison with Alternative VIX Option Pricing Models}
\label{sec:option_benchmarks}

We compare the following five models:
\begin{enumerate}
    \item \textbf{RISVIX}: The proposed rough ISVIX model, which introduces a fractional Volterra kernel into the dynamics of instantaneous squared VIX.
    
    \item \textbf{LZZ}: The Markovian ISVIX model of \citet{LZZ2019}, with $\alpha=1$.
    
    \item \textbf{HJT}: The modulated Volterra stochastic volatility model of \citet{HJT2020}, which introduces an additional stochastic factor to capture the VIX implied-volatility smile.
    
    \item \textbf{AIL}: The Gaussian polynomial volatility models of \citet{AbiJaberIllandLi2025}, in which VIX is derived from conditional integrated variance.
    
    \item \textbf{BPS}: The rough Hawkes--Heston model of \citet{BondiPulidoScotti2024}, which combines fractional volatility dynamics with self-exciting jumps and uses affine transform methods for VIX option pricing.
\end{enumerate}
RISVIX and LZZ model instantaneous squared VIX directly, whereas HJT, AIL and BPS derive VIX from the volatility dynamics of the underlying equity index.
\autoref{tab:option_benchmarks} reports their pricing errors on the same VIX call-option sample.

\begin{table}[!htbp]
\centering
\caption{\textbf{Comparison with alternative VIX option pricing models.}
\footnotesize{Calibration: January 2018--December 2023; evaluation: January 2024--June 2025.}}
\label{tab:option_benchmarks}
\small
\renewcommand{\arraystretch}{1.35}
\begin{tabular*}{\linewidth}{@{\extracolsep{\fill}}lcccc@{}}
\toprule
& \multicolumn{2}{c}{\textbf{Calibration}}
& \multicolumn{2}{c}{\textbf{Evaluation}} \\
\cmidrule(lr){2-3}
\cmidrule(l){4-5}
\textbf{Model} & \textbf{MSE} & \textbf{MAE} & \textbf{MSE} & \textbf{MAE} \\
\midrule
RISVIX
& \textbf{0.5015} & \textbf{0.5528}
& \textbf{0.6293} & \textbf{0.7061} \\
LZZ
& 0.5548 & 0.5815
& 0.7238 & 0.7537 \\
HJT
& 0.6497 & 0.6139
& 0.8070 & 0.7277 \\
AIL
& 0.6339 & 0.6521
& 0.8229 & 0.8265 \\
BPS
& 0.9218 & 0.6920
& 0.6757 & 0.7067 \\
\bottomrule
\end{tabular*}
\end{table}

RISVIX achieves the lowest MSE and MAE among the five models in both periods.
The relatively high calibration MSE of BPS mainly reflects larger pricing errors during the COVID-19 period in 2020.
In the evaluation period, BPS is the closest competitor: its MSE is 0.6757, compared with 0.6293 for RISVIX, while their MAEs are nearly identical at 0.7067 and 0.7061, respectively.

\section{Conclusion}
\label{sec5}

This paper develops RISVIX, a fractional extension of the ISVIX framework of \citet{LZZ2019} for modeling the VIX term structure and pricing VIX options.
We introduce a fractional Volterra kernel directly into instantaneous squared VIX and derive an explicit term structure governed by Mittag--Leffler decay, together with a semi-closed-form option-pricing formula based on an affine Fourier--Laplace transform.
The separation between term-structure and distributional parameters supports a two-step procedure that estimates memory, mean reversion and date-specific initial states from VIX indices before calibrating the remaining parameters to VIX option prices.

Using daily VIX term-structure indices and option quotes from January 2018 to June 2025, we find that RISVIX improves term-structure fit, particularly during periods of market stress.
Relative to LZZ, the full-sample MSE reductions are 10.42\%, 17.21\% and 13.18\% for VIX levels, slopes and curvature, respectively.
Over January--June 2020, these reductions increase to 17.60\%, 41.14\% and 46.59\%, with an estimated fractional parameter of $\widehat{\alpha}=0.7051$.

The option-pricing results show a similar pattern.
With global parameters fixed at their training-sample estimates and date-specific initial states updated from the contemporaneous VIX term structure, RISVIX reduces out-of-sample MSE and MAE by 13.06\% and 6.33\%, respectively, relative to LZZ.
The gains are larger for shorter-maturity options and high-VIX states, with an out-of-sample MSE reduction of 40.93\% when $\mathrm{VIX}_t\geq25$.
RISVIX also achieves lower aggregate errors than the additional term-structure and option-pricing benchmarks.

Future work could extend RISVIX to a joint SPX--VIX framework to impose consistency across equity and volatility markets \citep{CK2013,Romer2022}, or enrich the direct VIX specification with multiple memory factors and more general path-dependent feedback \citep{GuyonLekeufack2023,JacquierMuguruzaPannier2025}.

\section*{Acknowledgements}
Wen Su acknowledges the support from the EPSRC Centre for Doctoral Training in Mathematics of Random Systems: Analysis, Modelling and Simulation (EP/S023925/1).

\section*{Disclosure statement}
No potential conflict of interest was reported by the author(s).



\clearpage

\appendix

\counterwithin{equation}{section}

\renewcommand{\theequation}{\thesection.\arabic{equation}}
\renewcommand{\thetable}{\thesection.\arabic{table}}
\renewcommand{\thefigure}{\thesection.\arabic{figure}}
\renewcommand{\thelemma}{\thesection.\arabic{lemma}}

\setcounter{equation}{0}
\setcounter{table}{0}
\setcounter{figure}{0}
\setcounter{lemma}{0}
\setcounter{theorem}{0}
\setcounter{proposition}{0}
\setcounter{example}{0}

\section{Technical Lemmas}
\label{app:auxiliary_results}

\subsection{Resolvent Identities}
\label{app:resolvent_identities}

For scalar functions $f$ and $g$, write $(f*g)(t)=\int_0^t f(t-s)g(s)\,\mathrm{d}s$ whenever the integral exists.
For $\kappa>0$, the resolvent $R_\kappa$ of $\kappa K$ satisfies
\begin{equation}
\label{eq:app_resolvent_definition}
R_\kappa+\kappa K*R_\kappa=\kappa K,
\qquad
K-R_\kappa*K=\frac{1}{\kappa}R_\kappa.
\end{equation}
Consequently, for locally integrable $f$ and $g$, the equation $g=f-\kappa K*g$ is equivalent to $g=f-R_\kappa*f$.
Both implications follow by substitution and associativity of convolution.
This is the scalar resolvent identity underlying the variation-of-constants formula in \citet[Lemma~2.5]{ALP2019}.

For a locally square-integrable martingale $M$, write $(K*\mathrm{d}M)_t=\int_0^t K(t-s)\,\mathrm{d}M_s$.
When the convolutions are well defined and stochastic Fubini applies, the same calculation gives
\begin{equation}
\label{eq:app_stochastic_resolvent}
g=f-\kappa K*g+K*\mathrm{d}M
\quad\Longleftrightarrow\quad
g=f-R_\kappa*f+\frac{1}{\kappa}R_\kappa*\mathrm{d}M.
\end{equation}
Indeed, stochastic Fubini gives $R_\kappa*(K*\mathrm{d}M)=(R_\kappa*K)*\mathrm{d}M$, after which \autoref{eq:app_resolvent_definition} applies.
This argument also permits jumps in $M$; the identities are understood at times for which the indicated convolutions are defined.

\begin{lemma}
\label{lem:fractional_resolvent}
Let $K(t)=t^{\alpha-1}/\Gamma(\alpha)$ with $\alpha\in(1/2,1]$ and $\kappa>0$.
Then, for $t>0$,
\begin{equation}
\label{eq:app_fractional_resolvent}
\begin{aligned}
&R_\kappa(t)
=
\kappa t^{\alpha-1}E_{\alpha,\alpha}(-\kappa t^\alpha)
=
-\frac{\mathrm{d}}{\mathrm{d}t}E_{\alpha,1}(-\kappa t^\alpha),\\
&\int_0^t R_\kappa(s)\,\mathrm{d}s
=
1-E_{\alpha,1}(-\kappa t^\alpha),\\
&\int_0^t E_{\alpha,1}(-\kappa s^\alpha)\,\mathrm{d}s
=
tE_{\alpha,2}(-\kappa t^\alpha)
=
t\left[1-\kappa t^\alpha E_{\alpha,\alpha+2}(-\kappa t^\alpha)\right].
\end{aligned}
\end{equation}
And the weight $w$ in \autoref{eq:vix_term_structure} is strictly decreasing on $(0,\infty)$, with $w(0+)=1$ and $w(\infty)=0$.
\end{lemma}

\begin{proof}
The beta integral gives $K^{*n}(t)=t^{n\alpha-1}/\Gamma(n\alpha)$, where $K^{*n}$ denotes the $n$-fold convolution of $K$.
Hence the series
\begin{equation}
\label{eq:app_resolvent_series}
R_\kappa(t)
=
\sum_{n=1}^{\infty}(-1)^{n-1}\kappa^n K^{*n}(t)
=
\kappa t^{\alpha-1}E_{\alpha,\alpha}(-\kappa t^\alpha)
\end{equation}
converges in $L^1(0,T)$ for every finite $T$ and satisfies \autoref{eq:app_resolvent_definition} by cancellation of successive convolution powers.
Termwise differentiation of the Mittag--Leffler series on compact subsets of $(0,\infty)$ gives the derivative identity.
Integrating this identity and using $E_{\alpha,1}(0)=1$ gives the second line of \autoref{eq:app_fractional_resolvent}.
Termwise integration of the series for $E_{\alpha,1}$ gives the third line, with its second equality following from $E_{\alpha,2}(z)=1+zE_{\alpha,\alpha+2}(z)$.

For $0<\alpha<1$, the function $t\mapsto E_{\alpha,1}(-\kappa t^\alpha)$ is strictly decreasing from one to zero; see \citet{Mainardi2014}.
For $\alpha=1$, it equals $e^{-\kappa t}$.
Since $w(t)=t^{-1}\int_0^t E_{\alpha,1}(-\kappa s^\alpha)\,\mathrm{d}s$, its limits follow by continuity at zero and averaging at infinity.
Moreover, $w'(t)=[E_{\alpha,1}(-\kappa t^\alpha)-w(t)]/t<0$, which proves strict monotonicity.
\end{proof}

\noindent\textbf{Remark}.
Integrating the first identity in \autoref{eq:app_fractional_resolvent} over $[s,s+\tau_0]$ gives the averaged resolvent in \autoref{eq:vix_averaged_resolvent}.
This expression extends continuously to $s=0$; for $\alpha=1$, it reduces to $a_{\tau_0}(s)=w(\tau_0)e^{-\kappa s}$.

\subsection{Moment Conditions}
\label{app:risvix_admissibility}

Fix $T>0$, $\alpha\in(1/2,1)$, $\kappa,\mu_y>0$ and $\sigma_V,\sigma_\theta,\lambda^y\geq0$. Let $(V,\theta)$ be a nonnegative weak solution of \autoref{eq:rough_isvix_model} on $[0,T]$, with the driving processes, jump law and independence conditions specified immediately after that equation. The initial states are nonnegative and $\mathcal F_0$-measurable, with $\mathbb E^{\mathbb Q}[V_0^2+\theta_0^2]<\infty$. We require $\theta$ to be continuous and $V$ to be predictable with $\int_0^T V_s\,\mathrm{d}s<\infty$ almost surely; nonnegativity of $V$ and its Volterra equation are understood $\mathbb Q\otimes\mathrm{d}t$-almost everywhere.

\begin{lemma}
\label{lem:risvix_integrated_moments}
Under the preceding assumptions, $\theta$ is a square-integrable martingale and, for $0\leq s\leq t\leq T$, $\mathbb E_s^{\mathbb Q}[\theta_t]=\theta_s$ and $\mathbb E_s^{\mathbb Q}[\theta_t^2]=\theta_s^2+\sigma_\theta^2(t-s)\theta_s$ almost surely. Moreover,
\begin{equation}
\label{eq:app_risvix_integrated_moments}
\mathbb E^{\mathbb Q}\left[\int_0^T(V_t^2+\theta_t^2)\,\mathrm{d}t\right]
\leq C_T\left(1+\mathbb E^{\mathbb Q}[V_0^2+\theta_0^2]\right)<\infty,
\end{equation}
where $C_T$ depends only on $T,\alpha,\kappa,\sigma_V,\sigma_\theta,\lambda^y,\mu_y$. In particular, $\mathbb E_0^{\mathbb Q}[\int_0^T(V_t^2+\theta_t^2)\,\mathrm{d}t]<\infty$ almost surely.
\end{lemma}

\begin{proof}
Since $\theta$ is a nonnegative local martingale, it is a supermartingale and $\mathbb E^{\mathbb Q}[\int_0^T\theta_s\,\mathrm{d}s]\leq T\mathbb E^{\mathbb Q}[\theta_0]<\infty$. Thus its Brownian integral is square integrable, and so is $\theta$ because $\theta_0\in L^2(\mathbb Q)$. Conditional It\^o isometry and the martingale property give $\mathbb E_s^{\mathbb Q}[(\theta_t-\theta_s)^2]=\sigma_\theta^2(t-s)\theta_s$, proving the conditional moment identities. Consequently,
\begin{equation}
\mathbb E^{\mathbb Q}\left[\int_0^T\theta_t^2\,\mathrm{d}t\right]
=T\mathbb E^{\mathbb Q}[\theta_0^2]
+\frac{\sigma_\theta^2T^2}{2}\mathbb E^{\mathbb Q}[\theta_0].
\end{equation}

Let $R_\kappa$ be the resolvent in \autoref{eq:app_resolvent_definition} and set $q_\kappa(t)=E_{\alpha,1}(-\kappa t^\alpha)$. By \autoref{lem:fractional_resolvent}, $q_\kappa$ is bounded, $R_\kappa\in L^1(0,T)\cap L^2(0,T)$ and $K\in L^2(0,T)$. Define
\begin{equation}
\label{eq:app_risvix_localisation}
\tau_n=\inf\left\{t\in[0,T]:\int_0^t V_s\,\mathrm{d}s\geq n\right\}\wedge T,
\qquad I_s^{(n)}=\mathbf 1_{\{s\leq\tau_n\}}.
\end{equation}
The cumulative integral is continuous and adapted, so $\tau_n$ is a stopping time and $I^{(n)}$ is predictable. Pathwise integrability gives $\int_0^T I_s^{(n)}V_s\,\mathrm{d}s\leq n$ and $\tau_n\uparrow T$ almost surely. For any deterministic $h\in L^2(0,T)$, It\^o isometry and Tonelli's theorem yield
\begin{equation}
\label{eq:app_risvix_stopped_isometry}
\begin{aligned}
&\mathbb E^{\mathbb Q}\left[\int_0^T
\left|\int_0^t h(t-s)I_s^{(n)}\sqrt{V_s}\,\mathrm{d}B_{1,s}\right|^2
\mathrm{d}t\right]\\
&\qquad=
\mathbb E^{\mathbb Q}\left[\int_0^T\int_0^t
h(t-s)^2 I_s^{(n)}V_s\,\mathrm{d}s\,\mathrm{d}t\right]
\leq n\|h\|_{L^2(0,T)}^2.
\end{aligned}
\end{equation}
In particular, $r\mapsto(\int_0^r K(r-s)^2 I_s^{(n)}V_s\,\mathrm{d}s)^{1/2}$ belongs to $L^2(0,T)$ almost surely. Its convolution with $|R_\kappa|$ is therefore finite almost everywhere, verifying the stochastic-Fubini condition for the stopped Brownian term in \autoref{eq:app_stochastic_resolvent}. Ordinary Fubini applies pathwise to the drift and finite-activity jump terms. On $t<\tau_n$, the stopped Brownian integrals and their past time convolutions coincide with the unstopped ones. Thus \autoref{eq:app_stochastic_resolvent} and \autoref{eq:app_resolvent_definition} show that the process
\begin{equation}
\label{eq:app_risvix_localised_resolvent}
\begin{aligned}
V_t^{(n)}
={}&q_\kappa(t)V_0
+\int_0^t R_\kappa(t-s)\theta_s\,\mathrm{d}s\\
&+\frac{\sigma_V}{\kappa}
\int_0^t R_\kappa(t-s)I_s^{(n)}\sqrt{V_s}\,\mathrm{d}B_{1,s}
+\frac{1}{\kappa}\int_0^t R_\kappa(t-s)\,\mathrm{d}J_s^y
\end{aligned}
\end{equation}
agrees with $V_t$ on $\{t<\tau_n\}$, $\mathbb Q\otimes\mathrm{d}t$-almost everywhere.

Set $F_n=\mathbb E^{\mathbb Q}[\int_0^T|V_t^{(n)}|^2\,\mathrm{d}t]$. Young's convolution inequality, \autoref{eq:app_risvix_stopped_isometry} with $h=R_\kappa$, and the compensated-Poisson isometry with $\mathbb E^{\mathbb Q}[Y_1^2]=2\mu_y^2$ give
\begin{equation}
\label{eq:app_risvix_localised_bound}
\begin{aligned}
F_n\leq{}&
4T\|q_\kappa\|_{L^\infty(0,T)}^2\mathbb E^{\mathbb Q}[V_0^2]
+4\|R_\kappa\|_{L^1(0,T)}^2
\mathbb E^{\mathbb Q}\left[\int_0^T\theta_s^2\,\mathrm{d}s\right]\\
&+\frac{8\lambda^y\mu_y^2T}{\kappa^2}\|R_\kappa\|_{L^2(0,T)}^2
+\frac{4\sigma_V^2}{\kappa^2}\|R_\kappa\|_{L^2(0,T)}^2
\mathbb E^{\mathbb Q}\left[\int_0^T I_s^{(n)}V_s\,\mathrm{d}s\right].
\end{aligned}
\end{equation}
The right-hand side is finite for each $n$, since its last expectation is at most $n$; hence $F_n<\infty$ before any absorption argument. Agreement before $\tau_n$, nonnegativity of $V$, and Cauchy--Schwarz imply
\begin{equation}
\label{eq:app_risvix_first_moment_absorption}
\mathbb E^{\mathbb Q}\left[\int_0^T I_s^{(n)}V_s\,\mathrm{d}s\right]
\leq\mathbb E^{\mathbb Q}\left[\int_0^T|V_s^{(n)}|\,\mathrm{d}s\right]
\leq\sqrt{TF_n}.
\end{equation}
Substituting this bound into \autoref{eq:app_risvix_localised_bound}, using the moment formula for $\theta$ and $\mathbb E^{\mathbb Q}[\theta_0]\leq(1+\mathbb E^{\mathbb Q}[\theta_0^2])/2$, and applying Young's inequality give $F_n\leq C_T(1+\mathbb E^{\mathbb Q}[V_0^2+\theta_0^2])+F_n/2$, where $C_T$ depends only on the stated quantities and not on $n$. Since $\tau_n\uparrow T$, the agreement before $\tau_n$ gives $V^{(n)}\to V$ almost everywhere under $\mathbb Q\otimes\mathrm{d}t$. Fatou's lemma and the bound for $\theta$ prove \autoref{eq:app_risvix_integrated_moments}. Conditioning the resulting integrable random variable on $\mathcal F_0$ proves the final assertion.
\end{proof}

\begin{lemma}
\label{lem:risvix_exponential_moment}
Suppose that \autoref{eq:rough_isvix_model} admits the solution specified above on $[0,T]$.
Then there exists $c_T>0$, depending only on $T$ and the model parameters, such that
\begin{equation}
\label{eq:app_risvix_exponential_moment}
\mathbb E_0^{\mathbb Q}
\left[
\exp\left\{
c\int_0^T V_t\,\mathrm{d}t
\right\}
\right]
<\infty
\end{equation}
almost surely for every $c\in[0,c_T)$.
\end{lemma}

\begin{proof}
Let $X_T=\int_0^T V_t\,\mathrm{d}t$ and $q_\kappa(t)=E_{\alpha,1}(-\kappa t^\alpha)$. The resolvent calculation in the proof of \autoref{lem:risvix_integrated_moments}, followed by integration in time, gives
\begin{equation}
\label{eq:app_risvix_integrated_resolvent}
\begin{aligned}
X_T
={}&V_0\int_0^Tq_\kappa(t)\,\mathrm{d}t
+\int_0^T b_T(s)\theta_s\,\mathrm{d}s\\
&+\frac{\sigma_V}{\kappa}\int_0^T b_T(s)\sqrt{V_s}\,\mathrm{d}B_{1,s}
+\frac{1}{\kappa}\int_0^T b_T(s)\,\mathrm{d}J_s^y,
\end{aligned}
\end{equation}
where $b_T(s)=1-q_\kappa(T-s)\in[0,1]$ by \autoref{lem:fractional_resolvent}. Here $R_\kappa\in L^2(0,T)$ and $\mathbb E^{\mathbb Q}[X_T]<\infty$ justify stochastic Fubini for the Brownian term, while ordinary Fubini applies to the drift and finite-activity jump terms. Throughout the proof, $C_T$ denotes a positive deterministic constant depending only on $T$ and the model parameters, which may increase from line to line.

Choose $a>0$ such that $a\sigma_\theta^2T\leq1$. For fixed $s\in[0,T]$, applying It\^o's formula to $\exp\{a\theta_v/[1-a\sigma_\theta^2(s-v)/2]\}$, $0\leq v\leq s$, gives zero drift. Restricting first to $\{\theta_0\leq m\}$ and stopping this exponential at increasing levels of $\theta$, conditional Fatou's lemma, followed by $m\uparrow\infty$, yields
\begin{equation}
\mathbb E_0^{\mathbb Q}[e^{a\theta_s}]
\leq
\exp\left\{\frac{a\theta_0}{1-a\sigma_\theta^2s/2}\right\}
\leq e^{2a\theta_0}.
\end{equation}
Conditional Markov's inequality gives $\mathbb Q(\theta_s>2\theta_0+y\mid\mathcal F_0)\leq e^{-ay}$ for $y\geq0$. Integrating this tail bound shows that $(\mathbb E_0^{\mathbb Q}[\theta_s^p])^{1/p}\leq2\theta_0+(p!)^{1/p}/a\leq2\theta_0+p/a$ for every integer $p\geq2$. Conditional Minkowski's inequality then bounds the conditional $L^p$ norm of $\int_0^T b_T(s)\theta_s\,\mathrm{d}s$ by $C_T(\theta_0+p)$.

For the jump term, $|\int_0^T b_T(s)\,\mathrm{d}J_s^y|\leq\sum_{j=1}^{N_T^y}Y_j+\lambda^y\mu_yT$. Independence from $\mathcal F_0$ and the exponential jump law give $\mathbb E_0^{\mathbb Q}[\exp\{(2\mu_y)^{-1}\sum_{j=1}^{N_T^y}Y_j\}]=e^{\lambda^yT}$. Thus the last term in \autoref{eq:app_risvix_integrated_resolvent} has conditional $L^p$ norm at most $C_Tp$ for every integer $p\geq2$.

Set $x_p=(\mathbb E_0^{\mathbb Q}[X_T^p])^{1/p}$. By \autoref{lem:risvix_integrated_moments} and Cauchy--Schwarz, $x_2<\infty$ almost surely. The continuous-martingale Burkholder--Davis--Gundy inequality, with $L^p$ constant $2\sqrt{2p}$, and $b_T^2V\leq V$ give, for dyadic integers $p\geq4$ with $x_{p/2}<\infty$,
\begin{equation}
\label{eq:app_risvix_moment_recursion}
x_p
\leq
C_T(1+V_0+\theta_0+p)
+
C_T\sqrt p\,x_{p/2}^{1/2}.
\end{equation}
For the conditional inequality, first stop the Brownian integral and restrict to $\mathcal F_0$-measurable events on which $x_{p/2}$ is bounded; conditional Fatou removes these localisations. Thus the right-hand side is finite before any absorption argument, and dyadic induction gives $x_p<\infty$ for every dyadic $p\geq2$.

For dyadic $p\geq4$, use $x_{p/2}\leq x_p$ and Young's inequality in \autoref{eq:app_risvix_moment_recursion} to absorb half of $x_p$. Monotonicity of conditional $L^p$ norms between successive dyadic integers then gives $x_p\leq A_T+C_Tp$ for every integer $p\geq2$, where $A_T=C_T(1+V_0+\theta_0)<\infty$ almost surely and $C_T$ is deterministic. Hence $\mathbb E_0^{\mathbb Q}[X_T^n]\leq2^{n-1}(A_T^n+C_T^nn^n)$ for every integer $n\geq2$. Conditional monotone convergence and $n!\geq(n/e)^n$ in the exponential series show that $\mathbb E_0^{\mathbb Q}[e^{cX_T}]<\infty$ whenever $2eC_Tc<1$. Taking $c_T=(2eC_T)^{-1}$ proves \autoref{eq:app_risvix_exponential_moment}.
\end{proof}

\subsection{VIX Transform Regularity}

Fix $T,\tau_0>0$. For $\eta\geq0$, consider the real Riccati--Volterra equation
\begin{equation}
\label{eq:app_vix_real_riccati}
p_\eta(s)
=
\eta a_{\tau_0}(s)
+
\frac{\sigma_V^2}{2\kappa}\int_0^s R_\kappa(s-v)p_\eta(v)^2\,\mathrm{d}v,
\qquad 0\leq s\leq T,
\end{equation}
together with the Riccati equation
\begin{equation}
\label{eq:app_vix_real_second_riccati}
\psi_2'(s;\eta)
=
\kappa p_\eta(s)+\frac{\sigma_\theta^2}{2}\psi_2(s;\eta)^2,
\qquad
\psi_2(0;\eta)=\eta[1-w(\tau_0)].
\end{equation}

\begin{lemma}
\label{lem:vix_real_riccati}
There exists $\bar\eta_{T,\tau_0}>0$ such that, for every $\eta\in[0,\bar\eta_{T,\tau_0})$, the preceding equations admit a unique nonnegative continuous solution $(p_\eta,\psi_2(\cdot;\eta))$ on $[0,T]$, with
\begin{equation}
\label{eq:app_vix_real_riccati_bound}
\begin{aligned}
&\sup_{0\leq s\leq T}p_\eta(s)
\leq2\eta\|a_{\tau_0}\|_{L^\infty(0,T)}<\mu_y^{-1},\\
&\sup_{0\leq s\leq T}\psi_2(s;\eta)
\leq2\eta\left[1-w(\tau_0)+2\kappa T\|a_{\tau_0}\|_{L^\infty(0,T)}\right].
\end{aligned}
\end{equation}
In particular, both functions converge uniformly to zero as $\eta\downarrow0$.
\end{lemma}

\begin{proof}
By \autoref{lem:fractional_resolvent}, $R_\kappa$ is nonnegative and belongs to $L^1(0,T)$, $a_{\tau_0}$ is nonnegative and continuous on $[0,T]$, and $0<w(\tau_0)<1$. Let $A=\|a_{\tau_0}\|_{L^\infty(0,T)}$ and $L=\|R_\kappa\|_{L^1(0,T)}$. For $\eta>0$, define $\Gamma_\eta p=\eta a_{\tau_0}+(\sigma_V^2/(2\kappa))R_\kappa*p^2$. On the closed ball $\{p\in C([0,T];\mathbb R):\|p\|_\infty\leq2\eta A\}$,
\begin{equation}
\label{eq:app_vix_real_riccati_contraction}
\begin{aligned}
\|\Gamma_\eta p\|_\infty
\leq\eta A+\frac{2\sigma_V^2}{\kappa}L\eta^2A^2,\quad
\|\Gamma_\eta p-\Gamma_\eta q\|_\infty
\leq\frac{2\sigma_V^2}{\kappa}L\eta A\|p-q\|_\infty.
\end{aligned}
\end{equation}
For sufficiently small $\eta$, this map preserves the ball and is a contraction. Iteration from zero gives a nonnegative fixed point $p_\eta$. Reducing $\bar\eta_{T,\tau_0}$ if necessary ensures $2\eta A<\mu_y^{-1}$. If $\widetilde p_\eta$ is another nonnegative continuous solution, both are bounded and their difference is bounded in absolute value by $(\sigma_V^2/(2\kappa))R_\kappa*[(p_\eta+\widetilde p_\eta)|p_\eta-\widetilde p_\eta|]$. Since $\int_0^\delta R_\kappa(s)\,\mathrm{d}s\to0$ as $\delta\downarrow0$, contraction on successive intervals of a common sufficiently short length gives uniqueness on $[0,T]$.

Given $p_\eta$, the right-hand side of \autoref{eq:app_vix_real_second_riccati} is continuous in time and locally Lipschitz in $\psi_2$, so it has a unique maximal solution. Its derivative is nonnegative, and hence $\psi_2(s;\eta)\geq0$ on its interval of existence. Put $C=1-w(\tau_0)+2\kappa TA>0$ and reduce $\bar\eta_{T,\tau_0}$ further so that $2\sigma_\theta^2T\eta C<1$. If $\psi_2$ first reached $2\eta C$ at some $s\leq T$, its integral equation would give $\psi_2(s;\eta)\leq\eta C+2\sigma_\theta^2T\eta^2C^2<2\eta C$, a contradiction. Thus the solution and its derivative remain bounded. It has a finite limit at any endpoint up to $T$, and local existence prevents termination before $T$. It therefore extends to $[0,T]$ and satisfies the second bound in \autoref{eq:app_vix_real_riccati_bound}. At $\eta=0$, both solutions are identically zero, with uniqueness following from the same arguments. The two bounds prove uniform convergence as $\eta\downarrow0$.
\end{proof}

For $u\in\mathbb C$, consider
\begin{equation}
\label{eq:app_vix_complex_riccati}
\psi_1(s;u)
=
u a_{\tau_0}(s)
+
\frac{\sigma_V^2}{2\kappa}
\int_0^s
R_\kappa(s-v)\psi_1(v;u)^2\,\mathrm{d}v.
\end{equation}
Let $\eta\in[0,\bar\eta_{T,\tau_0})$ and $\xi\in\mathbb R$, and let $p_\eta$ be the solution supplied by \autoref{lem:vix_real_riccati}. Suppose that \autoref{eq:app_vix_complex_riccati} with $u=\eta+i\xi$ has a continuous solution $\psi_1(\,\cdot\,;\eta+i\xi)$ on $[0,T]$.

\begin{lemma}
\label{lem:vix_complex_comparison}
Under the preceding assumptions, this solution is unique and satisfies
\begin{equation}
\label{eq:app_vix_complex_comparison}
\operatorname{Re}\psi_1(s;\eta+i\xi)
\leq p_\eta(s)<\mu_y^{-1},
\qquad 0\leq s\leq T.
\end{equation}
\end{lemma}

\begin{proof}
Write $y=\operatorname{Im}\psi_1(\,\cdot\,;\eta+i\xi)$ and $d=p_\eta-\operatorname{Re}\psi_1(\,\cdot\,;\eta+i\xi)$. Subtracting the real part of \autoref{eq:app_vix_complex_riccati} from \autoref{eq:app_vix_real_riccati} gives
\begin{equation}
\label{eq:app_vix_comparison_difference}
d
=
\frac{\sigma_V^2}{2\kappa}R_\kappa*
\left[(2p_\eta-d)d+y^2\right].
\end{equation}
Since $p_\eta$ and $d$ are continuous on $[0,T]$, choose $\lambda\geq0$ such that $\lambda+(\sigma_V^2/(2\kappa))(2p_\eta-d)\geq0$ throughout this interval. The resolvent identity in \autoref{eq:app_resolvent_definition} gives $(I+\lambda R_\kappa*)^{-1}R_\kappa*=(1+\lambda)^{-1}R_{\kappa(1+\lambda)}*$. Adding $\lambda R_\kappa*d$ to both sides of \autoref{eq:app_vix_comparison_difference} therefore yields
\begin{equation}
\label{eq:app_vix_positive_comparison}
d
=
\frac{1}{1+\lambda}R_{\kappa(1+\lambda)}*
\left[
\left\{
\lambda+\frac{\sigma_V^2}{2\kappa}(2p_\eta-d)
\right\}d
+\frac{\sigma_V^2}{2\kappa}y^2
\right].
\end{equation}
Treat the coefficient in braces as the fixed continuous function determined by the given solution. The resulting linear Volterra equation has a nonnegative kernel and nonnegative forcing. Boundedness of this coefficient and $\int_0^\delta R_{\kappa(1+\lambda)}(s)\,\mathrm{d}s\to0$ as $\delta\downarrow0$ make Picard iteration contractive on sufficiently short intervals of a common length. The iterates from zero are nonnegative on the first interval; on each subsequent interval, the already determined history contributes another nonnegative forcing term. Uniqueness on each interval identifies these nonnegative solutions with $d$, so $d\geq0$ on $[0,T]$. Together with \autoref{eq:app_vix_real_riccati_bound}, this proves \autoref{eq:app_vix_complex_comparison}.

Finally, subtracting \autoref{eq:app_vix_complex_riccati} for two continuous solutions factors the difference of their squares into their sum times their difference. Both solutions are bounded and $R_\kappa\in L^1(0,T)$, so the same short-interval contraction argument proves uniqueness on $[0,T]$.
\end{proof}

\section{Proofs}

\subsection{Proof of \autoref{thm:isvix_conditional_mean}}
\label{app:proof_isvix_mean}

\begin{proof}
Fix $T>0$ and let $m(t)=\mathbb E_0^{\mathbb Q}[V_t]$ for Lebesgue-almost every $t\in[0,T]$, taking a jointly measurable version. 
By \autoref{lem:risvix_integrated_moments}, conditional Jensen's inequality and Fubini's theorem give $m\in L^2(0,T)$ almost surely. 
Since $K\in L^2(0,T)$ and $\mathbb E^{\mathbb Q}[\int_0^T V_s\,\mathrm{d}s]<\infty$, It\^o isometry shows that the Brownian convolution is square integrable for Lebesgue-almost every $t$. 
The compensated jump convolution is also square integrable by the compensated-Poisson isometry and $\mathbb E^{\mathbb Q}[Y_1^2]=2\mu_y^2$. 
Both convolutions therefore have zero $\mathcal F_0$-conditional mean.

Conditional Fubini and $\mathbb E_0^{\mathbb Q}[\theta_s]=\theta_0$ give, $\mathbb Q\otimes\mathrm{d}t$-almost everywhere,
\begin{equation}
\label{eq:app_risvix_mean_equation}
m(t)-\theta_0
=
V_0-\theta_0
-
\kappa\int_0^t K(t-s)\left[m(s)-\theta_0\right]\,\mathrm{d}s.
\end{equation}
Applying the resolvent identity in \autoref{eq:app_resolvent_definition} to this pathwise $L^1(0,T)$ equation yields, in the same almost-everywhere sense,
\begin{equation}
\label{eq:app_risvix_mean_solution}
m(t)-\theta_0
=
(V_0-\theta_0)
\left[
1-\int_0^t R_\kappa(s)\,\mathrm{d}s
\right]
=
(V_0-\theta_0)E_{\alpha,1}(-\kappa t^\alpha),
\end{equation}
where the second equality follows from \autoref{lem:fractional_resolvent}. Adding $\theta_0$ gives the right-hand side of \autoref{eq:isvix_conditional_mean}, which is continuous on $[0,T]$ and hence is the unique continuous representative of $m$.
\end{proof}

\subsection{Proof of \autoref{thm:vix_term_structure}}

\begin{proof}
Fix $t\in[0,T)$ and $\tau\in(0,T-t]$. Let $q_\kappa(s)=E_{\alpha,1}(-\kappa s^\alpha)$ and $M_u=\int_0^u\sigma_V\sqrt{V_s}\,\mathrm{d}B_{1,s}+J_u^y$. By \autoref{lem:risvix_integrated_moments}, both $\theta$ and $M$ are square-integrable martingales. The resolvent calculation in the proof of that lemma, using \autoref{eq:app_stochastic_resolvent}, gives, $\mathbb Q\otimes\mathrm{d}u$-almost everywhere on $[0,T]$,
\begin{equation}
\label{eq:app_vix_resolvent_representation}
V_u
=
q_\kappa(u)V_0
+
\int_0^u R_\kappa(u-s)\theta_s\,\mathrm{d}s
+
\frac{1}{\kappa}
\int_0^u R_\kappa(u-s)\,\mathrm{d}M_s.
\end{equation}

For the Brownian term, Tonelli's theorem and Cauchy--Schwarz give
\begin{equation}
\begin{aligned}
&\int_0^\tau
\left(
\mathbb E^{\mathbb Q}\left[
\int_0^{t+r}R_\kappa(t+r-s)^2V_s\,\mathrm{d}s
\right]
\right)^{1/2}\mathrm{d}r\\
&\qquad\leq
\sqrt{\tau}\,\|R_\kappa\|_{L^2(0,T)}
\left(
\mathbb E^{\mathbb Q}\left[\int_0^T V_s\,\mathrm{d}s\right]
\right)^{1/2}<\infty.
\end{aligned}
\end{equation}
This verifies the stochastic-Fubini condition for the window average. The finite-activity jump term admits pathwise Fubini, while $R_\kappa\in L^1(0,T)$ and the first moments supplied by \autoref{lem:risvix_integrated_moments} justify the deterministic and conditional Fubini interchanges. For Lebesgue-almost every $r\in(0,\tau)$, the stochastic integrals in \autoref{eq:app_vix_resolvent_representation} at $u=t+r$ are square integrable. Splitting them at $t$, using $\mathbb E_t^{\mathbb Q}[\theta_s]=\theta_t$ for $s\geq t$ and the zero conditional mean of the future martingale integrals, gives almost surely
\begin{equation}
\label{eq:app_vix_conditional_mean}
\begin{aligned}
\mathbb E_t^{\mathbb Q}[V_{t+r}]
={}&
q_\kappa(t+r)V_0
+
[1-q_\kappa(r)]\theta_t\\
&+
\int_0^t R_\kappa(t+r-s)\theta_s\,\mathrm{d}s
+
\frac{1}{\kappa}
\int_0^t R_\kappa(t+r-s)\,\mathrm{d}M_s,
\end{aligned}
\end{equation}
where \autoref{lem:fractional_resolvent} gives $\int_0^rR_\kappa(s)\,\mathrm{d}s=1-q_\kappa(r)$.

Averaging this almost-everywhere identity over $r\in(0,\tau)$ and applying conditional and stochastic Fubini yields
\begin{equation}
\label{eq:app_vix_average_intermediate}
\begin{aligned}
\mathrm{VIX}_{t,\tau}^{2}
={}&
\frac{V_0}{\tau}\int_t^{t+\tau}q_\kappa(s)\,\mathrm{d}s
+
[1-w(\tau)]\theta_t+
\int_0^t a_\tau(t-s)
\left\{
\kappa\theta_s\,\mathrm{d}s+\mathrm{d}M_s
\right\},
\end{aligned}
\end{equation}
because
\begin{equation}
\label{eq:app_vix_weights}
\begin{aligned}
&w(\tau)
=
\frac{1}{\tau}\int_0^\tau q_\kappa(s)\,\mathrm{d}s
=
E_{\alpha,2}(-\kappa\tau^\alpha),\\
&a_\tau(s)
=
\frac{1}{\kappa\tau}\int_s^{s+\tau}R_\kappa(v)\,\mathrm{d}v
=
\frac{q_\kappa(s)-q_\kappa(s+\tau)}{\kappa\tau}.
\end{aligned}
\end{equation}
Moreover, the second identity in \autoref{eq:app_vix_weights} gives
\begin{equation}
\label{eq:app_vix_initial_weight_identity}
\frac{1}{\tau}\int_t^{t+\tau}q_\kappa(s)\,\mathrm{d}s
=
w(\tau)
-
\kappa\int_0^t a_\tau(s)\,\mathrm{d}s.
\end{equation}
Since $a_\tau$ is bounded and continuous on $[0,t]$, the averaged martingale integral is square integrable for the fixed pair $(t,\tau)$. Substituting \autoref{eq:app_vix_initial_weight_identity} into \autoref{eq:app_vix_average_intermediate} and expanding $M$ therefore gives \autoref{eq:vix_term_structure} almost surely.
\end{proof}

\subsection{Proof of \autoref{thm:vix_transform}}

\begin{proof}
Fix a real parameter $u=\eta\in[0,\bar\eta_{T,\tau_0})$ and take $\psi_1=p_\eta$ and $\psi_2=\psi_2(\cdot;\eta)$ from \autoref{lem:vix_real_riccati}. Write $\mathrm{d}Z_s=\kappa(\theta_s-V_s)\,\mathrm{d}s+\sigma_V\sqrt{V_s}\,\mathrm{d}B_{1,s}+\mathrm{d}J_s^y$, so that $V-V_0=K*\mathrm{d}Z$ almost everywhere in time. By \autoref{eq:vix_term_structure},
\begin{equation}
\label{eq:app_vix_terminal_functional}
\begin{aligned}
\mathrm{VIX}_{T,\tau_0}^2
={}&w(\tau_0)V_0+[1-w(\tau_0)]\theta_T+\int_0^T\left(a_{\tau_0}+\kappa K*a_{\tau_0}\right)(T-s)\,\mathrm{d}Z_s.
\end{aligned}
\end{equation}
Here and below, boundedness of the deterministic coefficient functions, $K\in L^1(0,T)\cap L^2(0,T)$ and \autoref{lem:risvix_integrated_moments} justify the drift and Brownian Fubini interchanges; the finite-activity jump terms admit pathwise Fubini.

Applying $I+\kappa K*$ to \autoref{eq:app_vix_real_riccati} and using $R_\kappa+\kappa K*R_\kappa=\kappa K$ gives
\begin{equation}
\label{eq:app_vix_riccati_identity}
\psi_1
=u\left(a_{\tau_0}+\kappa K*a_{\tau_0}\right)
+K*\left(-\kappa\psi_1+\frac{\sigma_V^2}{2}\psi_1^2\right).
\end{equation}
Define the remaining coefficients by
\begin{equation}
\begin{aligned}
\chi_1(s)
&=u w(\tau_0)+\int_0^s\left[-\kappa\psi_1(v)+\frac{\sigma_V^2}{2}\psi_1(v)^2\right]\mathrm{d}v,\\
\phi(s)
&=\lambda^y\int_0^s\left[\varrho(\psi_1(v))-1-\mu_y\psi_1(v)\right]\mathrm{d}v.
\end{aligned}
\end{equation}
Substituting \autoref{eq:app_vix_riccati_identity} into \autoref{eq:app_vix_terminal_functional} and using $V-V_0=K*\mathrm{d}Z$ under stochastic Fubini yields
\begin{equation}
\begin{aligned}
u\mathrm{VIX}_{T,\tau_0}^2
={}&\chi_1(T)V_0+u[1-w(\tau_0)]\theta_T
+\kappa\int_0^T\psi_1(T-s)\theta_s\,\mathrm{d}s\\
&+\int_0^T\sigma_V\psi_1(T-s)\sqrt{V_s}\,\mathrm{d}B_{1,s}
+\int_0^T\psi_1(T-s)\,\mathrm{d}J_s^y\\
&-\frac{\sigma_V^2}{2}\int_0^T\psi_1(T-s)^2V_s\,\mathrm{d}s.
\end{aligned}
\end{equation}

For the mean factor, apply It\^o's formula to $\psi_2(T-s)\theta_s$. Since $\mathrm{d}\theta_s=\sigma_\theta\sqrt{\theta_s}\,\mathrm{d}B_{2,s}$ and $\psi_2$ satisfies \autoref{eq:app_vix_real_second_riccati},
\begin{equation}
\begin{aligned}
&u[1-w(\tau_0)]\theta_T+\kappa\int_0^T\psi_1(T-s)\theta_s\,\mathrm{d}s\\
&\qquad=\psi_2(T)\theta_0
+\int_0^T\sigma_\theta\psi_2(T-s)\sqrt{\theta_s}\,\mathrm{d}B_{2,s}
-\frac{\sigma_\theta^2}{2}\int_0^T\psi_2(T-s)^2\theta_s\,\mathrm{d}s.
\end{aligned}
\end{equation}
Combining these identities and exponentiating gives
\begin{equation}
\label{eq:app_vix_transform_factorization}
\exp\left\{u\mathrm{VIX}_{T,\tau_0}^2\right\}
=\exp\left\{\phi(T)+\chi_1(T)V_0+\psi_2(T)\theta_0\right\}\mathcal E_T,
\end{equation}
where, for $0\leq t\leq T$,
\begin{equation}
\label{eq:app_vix_stochastic_exponential}
\begin{aligned}
\mathcal E_t=\exp\Bigg\{&
\int_0^t\sigma_V\psi_1(T-s)\sqrt{V_s}\,\mathrm{d}B_{1,s}
+\int_0^t\sigma_\theta\psi_2(T-s)\sqrt{\theta_s}\,\mathrm{d}B_{2,s}\\
&-\frac12\int_0^t\left[\sigma_V^2\psi_1(T-s)^2V_s+\sigma_\theta^2\psi_2(T-s)^2\theta_s\right]\mathrm{d}s\\
&+\int_0^t\psi_1(T-s)\,\mathrm{d}J_s^y
-\lambda^y\int_0^t\left[\varrho(\psi_1(T-s))-1-\mu_y\psi_1(T-s)\right]\mathrm{d}s
\Bigg\}.
\end{aligned}
\end{equation}

We first prove the martingale property for real $u=\eta\in[0,\bar\eta_{T,\tau_0})$, taking $\psi_1=p_\eta$ and $\psi_2=\psi_2(\cdot;\eta)$ from \autoref{lem:vix_real_riccati}. Write $\mathcal E_t=D_t\exp\{L_t-\langle L\rangle_t/2\}$, where $L$ is the sum of the two Brownian integrals and $D$ is the jump exponential in \autoref{eq:app_vix_stochastic_exponential}. The continuous and jump factors have zero quadratic covariation, so $\mathcal E$ is a nonnegative local martingale. Since $B_1$ and $B_2$ are independent, $\langle L\rangle_T=\int_0^T[\sigma_V^2p_\eta(T-s)^2V_s+\sigma_\theta^2\psi_2(T-s;\eta)^2\theta_s]\,\mathrm{d}s$.

Choose $a>0$ with $a\sigma_\theta^2T\leq1$. The bound for $\mathbb E_0^{\mathbb Q}[e^{a\theta_s}]$ in the proof of \autoref{lem:risvix_exponential_moment} and Jensen's inequality give $\mathbb E_0^{\mathbb Q}[\exp\{(a/T)\int_0^T\theta_s\,\mathrm{d}s\}]\leq e^{2a\theta_0}$. Both coefficient functions converge uniformly to zero as $\eta\downarrow0$, so reduce $\bar\eta_{T,\tau_0}$ until $2\sigma_V^2\|p_\eta\|_\infty^2<c_T$, $2\sigma_\theta^2\|\psi_2(\cdot;\eta)\|_\infty^2<a/T$ and $2\|p_\eta\|_\infty<\mu_y^{-1}$ throughout this interval, with norms on $[0,T]$. Conditional Cauchy--Schwarz and \autoref{lem:risvix_exponential_moment} then yield $\mathbb E_0^{\mathbb Q}[e^{\langle L\rangle_T}]<\infty$ almost surely.

The exponential formula for the marked Poisson process shows that $D$ is a strictly positive martingale with $\mathbb E_0^{\mathbb Q}[D_T]=1$ and
\begin{equation}
\mathbb E_0^{\mathbb Q}[D_T^2]
=\exp\left\{\lambda^y\int_0^T\left[\varrho(2p_\eta(s))-2\varrho(p_\eta(s))+1\right]\mathrm{d}s\right\}<\infty.
\end{equation}
Define $\mathbb Q^y$ on $\mathcal F_T$ by $\mathrm{d}\mathbb Q^y/\mathrm{d}\mathbb Q=D_T$. This change of measure leaves $B_1,B_2$ as a two-dimensional Brownian motion: the density martingale has zero quadratic covariation with each Brownian motion, so Girsanov's theorem introduces no continuous drift. Moreover, $\mathbb Q^y$ agrees with $\mathbb Q$ on $\mathcal F_0$, and conditional Cauchy--Schwarz gives
\begin{equation}
\label{eq:app_vix_novikov}
\begin{aligned}
\mathbb E_0^{\mathbb Q^y}[e^{\langle L\rangle_T/2}]
=\mathbb E_0^{\mathbb Q}[D_Te^{\langle L\rangle_T/2}]\leq\left(\mathbb E_0^{\mathbb Q}[D_T^2]\right)^{1/2}
\left(\mathbb E_0^{\mathbb Q}[e^{\langle L\rangle_T}]\right)^{1/2}<\infty.
\end{aligned}
\end{equation}
To apply Novikov's criterion, first condition $\mathbb Q^y$ on the $\mathcal F_0$-measurable events $\{\mathbb E_0^{\mathbb Q^y}[e^{\langle L\rangle_T/2}]\leq n\}$ of positive probability. The Brownian property is preserved and the required unconditional expectation is then finite. Novikov's criterion applies to the joint continuous exponential under each such conditional measure. These events increase to the whole space up to a null set, giving $\mathbb E_0^{\mathbb Q^y}[\exp\{L_T-\langle L\rangle_T/2\}]=1$. Bayes' formula therefore yields $\mathbb E_0^{\mathbb Q}[\mathcal E_T]=1$, so $\mathcal E$ is a uniformly integrable martingale on $[0,T]$. 
When $\lambda^y=0$, take $D\equiv1$ and $\mathbb Q^y=\mathbb Q$; the same argument applies.
Taking conditional expectations in \autoref{eq:app_vix_transform_factorization} therefore gives $f(\eta;0,T)=\exp\{\phi(T)+\chi_1(T)V_0+\psi_2(T)\theta_0\}<\infty$ almost surely for every sufficiently small real $\eta\geq0$, with the coefficients specified above.

It remains to extend the identity to complex arguments.
By the definition of squared VIX, conditional Jensen's inequality, and
\autoref{lem:risvix_exponential_moment}, after reducing
$\bar\eta_{T,\tau_0}$ once more if necessary,
\begin{equation}
\label{eq:app_vix_transform_moment_strip}
\begin{aligned}
\mathbb E_0^{\mathbb Q}
\left[
e^{\eta\mathrm{VIX}_{T,\tau_0}^2}
\right]
&\leq
\mathbb E_0^{\mathbb Q}
\left[
\exp\left\{
\frac{\eta}{\tau_0}
\int_T^{T+\tau_0}V_s\,\mathrm{d}s
\right\}
\right]\\
&\leq
\mathbb E_0^{\mathbb Q}
\left[
\exp\left\{
\frac{\eta}{\tau_0}
\int_0^{T+\tau_0}V_s\,\mathrm{d}s
\right\}
\right]
<\infty
\end{aligned}
\end{equation}
for every $\eta\in(0,\bar\eta_{T,\tau_0})$.
Fix $u=\eta+i\xi$ with $0\leq\eta<\bar\eta_{T,\tau_0}$ and $\xi\in\mathbb R$. Since $R_\kappa\in L^1(0,T)$ and the nonlinearity is polynomial, contraction on short intervals gives a unique maximal continuous solution of \autoref{eq:app_vix_complex_riccati}. The comparison argument in \autoref{lem:vix_complex_comparison} applies on every compact interval of existence and gives $\operatorname{Re}\psi_1(s;u)\leq p_\eta(s)$.

To exclude explosion, write $D_C^\alpha g(s)=\Gamma(1-\alpha)^{-1}\int_0^s(s-v)^{-\alpha}g'(v)\,\mathrm{d}v$ for the Caputo derivative of an absolutely continuous function. The resolvent identity rewrites the equation as $\psi_1=u a_{\tau_0}+K*[-\kappa\psi_1+\kappa u a_{\tau_0}+(\sigma_V^2/2)\psi_1^2]$. On each compact interval of existence, boundedness first gives $\alpha$-H\"older continuity of $\psi_1$ and of the integrand. Since $2\alpha>1$, differentiating the convolution of this integrand with $K$ gives an integrable bound of the form $C(s^{\alpha-1}+s^{2\alpha-1})$ near zero. Together with absolute continuity of $a_{\tau_0}$, this proves that $\psi_1$ is absolutely continuous on that interval. Its Caputo equation is therefore
\begin{equation}
\begin{aligned}
D_C^\alpha\psi_1(s;u)
&=u h(s)-\kappa\psi_1(s;u)+\frac{\sigma_V^2}{2}\psi_1(s;u)^2,
\qquad \psi_1(0;u)=u a_{\tau_0}(0),\\
h(s)
&=-\frac{E_{\alpha,1}(-\kappa(s+\tau_0)^\alpha)}{\tau_0}
+\frac{1}{\kappa\tau_0\Gamma(1-\alpha)}
\int_0^s(s-v)^{-\alpha}R_\kappa(v+\tau_0)\,\mathrm{d}v.
\end{aligned}
\end{equation}
Here $h=D_C^\alpha a_{\tau_0}+\kappa a_{\tau_0}$, and its displayed expression follows from \autoref{lem:fractional_resolvent}. Since $\tau_0>0$, the function $R_\kappa(\cdot+\tau_0)$ is bounded on $[0,T]$, so $h$ is bounded as well.

Fractional integration by parts gives $D_C^\alpha|\psi_1|^2\leq2\operatorname{Re}(\overline{\psi_1}D_C^\alpha\psi_1)$. Indeed, the difference between the right-hand side and the left-hand side equals
\begin{equation}
\frac{1}{\Gamma(1-\alpha)}
\left[
\frac{|\psi_1(s;u)-\psi_1(0;u)|^2}{s^\alpha}
+\alpha\int_0^s
\frac{|\psi_1(s;u)-\psi_1(v;u)|^2}{(s-v)^{1+\alpha}}\,\mathrm{d}v
\right]\geq0.
\end{equation}
Consequently, the real-part comparison yields
\begin{equation}
\begin{aligned}
D_C^\alpha|\psi_1(s;u)|^2
&\leq
2\operatorname{Re}\!\left(\overline{\psi_1(s;u)}\,u h(s)\right)
+\left[-2\kappa+\sigma_V^2\operatorname{Re}\psi_1(s;u)\right]|\psi_1(s;u)|^2\\
&\leq
|u|^2\|h\|_\infty^2
+\left(1+\sigma_V^2\|p_\eta\|_\infty\right)|\psi_1(s;u)|^2.
\end{aligned}
\end{equation}
Applying $K*$ and iterating the resulting linear inequality gives
\begin{equation}
|\psi_1(s;u)|^2
\leq
|u|^2
\left[
a_{\tau_0}(0)^2+\frac{\|h\|_\infty^2T^\alpha}{\Gamma(1+\alpha)}
\right]
E_{\alpha,1}\!\left(
[1+\sigma_V^2\|p_\eta\|_\infty]T^\alpha
\right)
\end{equation}
throughout its interval of existence before $T$. This bound is independent of the endpoint of that interval. Boundedness of $\psi_1^2$ and translation continuity of $R_\kappa$ in $L^1$ give a continuous limit of the convolution at any finite endpoint. The resulting continuous history term permits another short-interval contraction. Thus $\psi_1$ extends uniquely to $[0,T]$, with $\operatorname{Re}\psi_1(s;u)\leq p_\eta(s)<\mu_y^{-1}$.

Given this solution, consider the complex Riccati equation
\begin{equation}
\psi_2'(s;u)
=\kappa\psi_1(s;u)+\frac{\sigma_\theta^2}{2}\psi_2(s;u)^2,
\qquad
\psi_2(0;u)=u[1-w(\tau_0)].
\end{equation}
Its right-hand side is locally Lipschitz, so it has a unique maximal solution. Subtracting its real part from \autoref{eq:app_vix_real_second_riccati} gives a linear equation for $\psi_2(s;\eta)-\operatorname{Re}\psi_2(s;u)$, with zero initial value and nonnegative forcing $\kappa[p_\eta(s)-\operatorname{Re}\psi_1(s;u)]+(\sigma_\theta^2/2)[\operatorname{Im}\psi_2(s;u)]^2$. An integrating factor therefore gives $\operatorname{Re}\psi_2(s;u)\leq\psi_2(s;\eta)$ before any possible explosion. Hence
\begin{equation}
\begin{aligned}
\frac{\mathrm{d}}{\mathrm{d}s}|\psi_2(s;u)|^2
&=
2\kappa\operatorname{Re}\!\left(\overline{\psi_2(s;u)}\psi_1(s;u)\right)
+\sigma_\theta^2\operatorname{Re}\psi_2(s;u)|\psi_2(s;u)|^2\\
&\leq
\kappa|\psi_1(s;u)|^2
+\left[\kappa+\sigma_\theta^2\|\psi_2(\cdot;\eta)\|_\infty\right]|\psi_2(s;u)|^2.
\end{aligned}
\end{equation}
The preceding bound for $\psi_1$ and Gronwall's inequality exclude explosion of $\psi_2$ before $T$. Thus both coefficient functions exist uniquely on $[0,T]$ throughout the stated strip.

These bounds are uniform when $u$ ranges over a compact subset of the strip. Picard iteration on successive sufficiently short intervals therefore converges locally uniformly in $u$. Since the integral equations are polynomial in their unknown functions and in $u$, their iterates are holomorphic, proving that $u\mapsto\psi_1(\cdot;u)$ and $u\mapsto\psi_2(\cdot;u)$ are holomorphic with values in $C([0,T])$ on $0<\operatorname{Re}(u)<\bar\eta_{T,\tau_0}$. The same construction gives continuous dependence at $\operatorname{Re}(u)=0$. Define $\chi_1$ and $\phi$ by the integrals in the real-parameter part of the proof. Their holomorphic dependence follows from that of $\psi_1$ and the bound $\operatorname{Re}\psi_1<\mu_y^{-1}$.

Finally, take the conditional moment bounds in \autoref{eq:app_vix_transform_moment_strip} and the real-parameter transform identities simultaneously on a countable dense subset of $(0,\bar\eta_{T,\tau_0})$. Outside a common null set, a conditional distribution of $\mathrm{VIX}_{T,\tau_0}^2$ defines a holomorphic version of $f(u;0,T)$ on the open strip: on each compact subset, an exponential moment at a larger real parameter dominates the integrand and its derivatives. The affine expression is holomorphic on the same strip and agrees with this version on the chosen dense set. The identity theorem therefore proves \autoref{eq:vix_transform} throughout the open strip, with the coefficients specified above. Conditional dominated convergence and continuity of the coefficient functions extend the identity to $\operatorname{Re}(u)=0$.
\end{proof}

\subsection{Proof of \autoref{thm:vix_call_price}}

\begin{proof}
For $\operatorname{Re}(z)>0$, the Laplace transform of the payoff
$x\mapsto(\sqrt{x}-K)^+$ is
\begin{equation}
\label{eq:app_vix_payoff_laplace}
\int_0^\infty
e^{-zx}(\sqrt{x}-K)^+\,\mathrm{d}x
=
\frac{\sqrt{\pi}}{2z^{3/2}}
\operatorname{erfc}(K\sqrt z).
\end{equation}
Indeed, this follows by substituting $x=y^2$ and integrating the resulting
Gaussian terms.

Fix $\eta\in(0,\bar\eta_{T,\tau_0})$. By \autoref{thm:vix_transform}, the entire contour $u=\eta+i\xi$, $\xi\in\mathbb R$, lies in the transform domain, with $f(\eta;0,T)<\infty$ and $|f(\eta+i\xi;0,T)|\leq f(\eta;0,T)$ almost surely. For $z=\eta+i\xi$, two integrations by parts in \autoref{eq:app_vix_payoff_laplace} give
\begin{equation}
\begin{aligned}
\frac{\sqrt{\pi}}{2z^{3/2}}\operatorname{erfc}(K\sqrt z)
&=\frac{e^{-zK^2}}{2Kz^2}
-\frac{1}{4z^2}\int_{K^2}^{\infty}e^{-zx}x^{-3/2}\,\mathrm{d}x,\\
\left|\frac{\sqrt{\pi}}{2z^{3/2}}\operatorname{erfc}(K\sqrt z)\right|
&\leq\frac{e^{-\eta K^2}}{K(\eta^2+\xi^2)}.
\end{aligned}
\end{equation}
The bound uses $K>0$ and $\int_{K^2}^{\infty}x^{-3/2}\,\mathrm{d}x=2/K$, and proves integrability of the payoff transform along the contour.

The function $x\mapsto e^{-\eta x}(\sqrt{x}-K)^+$ on $[0,\infty)$, extended by zero to negative $x$, is continuous and integrable. Fourier inversion therefore gives its representation at every $x$, and multiplication by $e^{\eta x}$ yields the Bromwich representation of the payoff. To interchange this integral with conditional expectation, restrict first to the $\mathcal F_0$-measurable events $\{f(\eta;0,T)\leq n\}$. After multiplication by the corresponding indicator, the expected integral of the absolute inversion integrand is bounded by $n\pi e^{-\eta K^2}/(K\eta)<\infty$. Ordinary Fubini applies on these events, which increase to the whole space up to a null set. Thus Bromwich inversion and conditional Fubini give
\begin{equation}
\label{eq:app_vix_payoff_inversion}
\begin{aligned}
\mathbb E_0^{\mathbb Q}
\left[
(\mathrm{VIX}_{T,\tau_0}-K)^+
\right]
&=
\frac{1}{4\sqrt{\pi}}
\int_{-\infty}^{\infty}
\frac{
\operatorname{erfc}\!\left(K\sqrt{\eta+i\xi}\right)
}{
(\eta+i\xi)^{3/2}
}
f(\eta+i\xi;0,T)\,\mathrm{d}\xi\\
&=
\frac{1}{2\sqrt{\pi}}
\int_0^\infty
\operatorname{Re}
\left[
\frac{
\operatorname{erfc}\!\left(K\sqrt{\eta+i\xi}\right)
}{
(\eta+i\xi)^{3/2}
}
f(\eta+i\xi;0,T)
\right]\mathrm{d}\xi.
\end{aligned}
\end{equation}
The second equality follows from conjugate symmetry.
Multiplying by $e^{-rT}$ gives \autoref{eq:vix_call_price}.
\end{proof}

\end{document}